\documentstyle[12pt,epsfig]{article}
\newcommand{\bea}{\begin{eqnarray}}
\newcommand{\ena}{\end{eqnarray}}
\newcommand{\vs}[1]{\vspace{#1 mm}}

\newcommand{\PL}[1]{Phys.\ Lett.\ {\bf #1}}

\newcommand{\EPJ}[1]{Eur.\ Phys.\ J.\ {\bf #1}}
\newcommand{\tp}{{\tilde {z}}}

\begin{document}
\noindent
\topmargin 0pt
\oddsidemargin 5mm

\setcounter{page}{0}
\thispagestyle{empty}
\begin{flushright}
February 9, 2001\\
OU-HET 376\\
hep-ph/0102111\\
\end{flushright}
\vs{4}
\begin{center}
{\LARGE{\bf  The matter effect to T-violation}} \\ 
\vs{2}
{\LARGE{\bf at a neutrino factory}}\\
\vs{6}
{\large 
Takahiro Miura\footnote{e-mail address:
miura@het.phys.sci.osaka-u.ac.jp},
Eiichi Takasugi\footnote{e-mail address:
takasugi@het.phys.sci.osaka-u.ac.jp},
Yoshitaka Kuno\footnote{e-mail address:
kuno@phys.sci.osaka-u.ac.jp}
\\
\vs{2}
{\em Department of Physics,
Osaka University \\ Toyonaka, Osaka 560-0043, Japan} \\
\vs{2}
Masaki Yoshimura\footnote{e-mail address:
myv20012@se.ritsumei.ac.jp}\\
\vs{2}
{\em Department of Physics,
Ritsumeikan University \\ Kusatsu, Shiga 525-8577, Japan} }
\end{center}
\vs{6}
\centerline{{\bf Abstract}}
We analyzed T-violation in neutrino oscillation by using 
perturbation methods with respect to $\Delta m_{21}^2L/2E$ 
and $\delta a(x)L/2E$, where $\delta a(x)$ represents the matter 
density fluctuation from its average value.  We found that the matter 
contribution to T-violation arises from interferences 
between $\Delta m_{21}^2L/2E$ and $\delta a(x) L/2E$. In the 
2nd order, the symmetric and asymmetric matter density fluctuations 
give effects to the $\sin \delta$ (intrinsic) 
and the $\cos \delta$ (fake) parts of 
T-violation. We give their analytic forms and 
analyze the matter contribution to the $\sin \delta$ and $\cos
\delta$ terms. 
We found that, for $L=3000$km, both the symmetric and asymmetric 
matter density fluctuations give negligible contributions to 
T-violation, and that thus the constant (average) matter density gives a 
good approximation. On the other hand, we argue that,  
for $L=7000$km or longer length, T-violation turns out to become 
very small due to cancellation between the 1st and the 2nd order 
terms. This shows that the constant (average) matter approximation 
is not valid. 

\vskip 1cm
\section{Introduction}
 
A high-intensity neutrino source based on a muon storage ring, which 
is now generally called a neutrino factory[1], has attracted growing 
interest from theorists and experimentalists[2]. One of the 
important physics potentials at neutrino factories is to measure a 
possible non-zero CP violation phase ($\delta$) in the 3-generation 
neutrino mixing matrix, the Maki-Nakagawa-Sakata (MNS) matrix[3]. 
To measure the phase $\delta$ in the neutrino 
oscillation, one way is to compare the CP-conjugate oscillation 
processes, $P(\nu_{\alpha}\rightarrow \nu_{\beta})$ and 
$P(\overline{\nu}_{\alpha} \rightarrow \overline{\nu}_{\beta})$, and 
the other way is to compare the T-reversed oscillation processes, 
$P(\nu_{\alpha} \rightarrow \nu_{\beta})$ and $P(\nu_{\beta} 
\rightarrow \nu_{\alpha})$. The study of CP violation has 
been extensively studied[4]. 

T-violation  has not been discussed seriously so far. 
It is mostly due to the difficulty of identification of 
$\nu_e$ appearance at a neutrino factory, 
since the detection of wrong-signed electrons is hard.
However, there have been some attempts 
to disentangle wrong-signed 
electrons which enables us to search for T-violation at a neutrino 
factory[1]. Its experimental feasibility studies are now undertaken. 
One of the advantages of the search for T-violation is to expect 
relatively small contribution from matter, whereas 
in the search for CP violation 
the fake CP-odd effects from matter dominates over the 
intrinsic CP violation for a long baseline length (such as more 
than a few thousand km) and thereby the measurement of the intrinsic
CP violation becomes challenging. 

T-violation  in matter arises from the intrinsic 
contribution with matter modification, which is proportional to 
the CP phase, and the fake matter contribution.  
The constant matter density gives an effect to the intrinsic 
T-violation which is proportional to
the $\sin \delta$ term, which has been discussed 
extensively[5, 6, 7].
However, up to now, it has not been addressed 
about possible contributions from the symmetric   
and also asymmetric matter fluctuations deviated from 
the average constant density. 

In this paper, we consider the symmetric and asymmetric matter 
fluctuations from the average density and treat them 
in the perturbation method developed by Koike and Sato[8], 
and Ota and Sato[9], where the quantities,  
$\Delta m_{21}^2L/2E$ and $\delta a(x)L/2E$, are considered as 
perturbative Hamiltonians, which may be small for most of the cases. 
The average matter density is included in the 
unperturbed Hamiltonian, so that the constant matter 
contribution (the average matter) was taken into account. 
The matter fluctuation is separated into the symmetric and 
asymmetric terms, $\delta a(x)_sL/2E$ and 
$\delta a(x)_aL/2E$, respectively. 
Ota and Sato considered the symmetric part by using the 
preliminary reference earth model (PREM)[10] and 
analyzed the 1st order corrections for $\Delta m_{21}^2L/2E$ 
and $\delta a(x)_sL/2E$. In the 1st order, T-violation 
arises only through $\Delta m_{21}^2L/2E$. The 
matter fluctuation does not give any contribution 
to T-violation in this order.  

We examined T-violation in the 2nd order 
perturbation of $\Delta m_{21}^2L/2E$ and $\delta a(x)L/2E$. 
Our motivation is to obtain the analytic expression of 
T-violation  in the 2nd order and to examine 
how large the 2nd order contribution from the symmetric and 
asymmetric fluctuations is. 
We found that the interference term between 
$\Delta m_{21}^2L/2E$ and $\delta a(x) L/2E$ gives some 
contributions to T-violation. In particular, 
the symmetric matter fluctuation contributes to 
the $\sin \delta$ part, while the asymmetric matter fluctuation 
does to the $\cos \delta$ part. We estimated 
these contributions for $L=3000$km case and found that 
these contributions are negligibly small for most energies. 
As a result, the constant matter approximation works well. 
On the other hand, for $L=7000$km or longer, the 2nd order 
term becomes comparable or larger than the 1st order term, 
so that the matter fluctuation can not be neglected and 
the constant matter density approximation fails.

In Sec.2, the perturbation formula is given, and the 0th and the 
1st order contributions with respect to $\Delta m_{21}^2L/2E$ are 
given in Sec.3. The general discussion on the contribution from 
the asymmetric matter profile to T-violation  is given in Sec.4. 
In Sec.5, the 2nd order contribution from $\Delta m_{21}^2L/2E$ is 
given. The interference term between $\Delta m_{21}^2L/2E$ and 
$\delta a(x)L/2E$ is presented in Sec.6 by assuming the linear 
dependence for $\delta a(x)$ and the numerical analysis of these 
interference terms is given in Sec.7.  The summary is given in 
Sec.8.

\section{The perturbation formula} 

The formula to evaluate the neutrino transition probabilities 
perturbatively with respect to the small quantities, 
$\Delta m_{21}^2L/2E$ and $\delta a(x)L/2E$, has been developed by 
Koike and Sato[8], and Ota and 
Sato[9]. Ota and Sato used this formula to estimated the 1st 
order terms of $\Delta m_{21}^2L/2E$ and the symmetric matter 
fluctuation, $\delta a(x)_s L/2E$. Here, we 
calculate the higher order terms with respect to 
the symmetric, $\delta a(x)_s L/2E$, and the asymmetric terms, 
$\delta a(x)_a L/2E$. Firstly, we outline their method.

We begin with defining the neutrino mixing matrix as
\bea
U
&=&e^{i\theta_{y} \lambda_7}{\rm diag }(1,1,e^{i\delta})
 e^{i\theta_{z} \lambda_5}e^{i\theta_{x}\lambda_2}
\nonumber\\
&=&\pmatrix{
c_{x} c_{z} &s_{x} c_{z}  & s_{z} \cr
-s_{x} c_{y} -c_{x} s_{y} s_{z} e^{i\delta} &
c_{x} c_{y} -s_{x} s_{y} s_{z} e^{i\delta} &
s_{y} c_{z} e^{i\delta}\cr 
s_{x} s_{y}-c_{x} c_{y} s_{z} e^{i\delta} &
-c_{x} s_{y}-s_{x} c_{y} s_{z} e^{i\delta} &
c_{y} c_{z}e^{i\delta}\cr}\;,
\ena
where $\lambda_j$ ($j=2,5,7$) are Gell-Mann matrices and 
$c_{a}=\cos \theta_{a}$ and $s_{a}=\sin \theta_{a}$. 
The angles $\theta_x$, $\theta_y$ and 
$\theta_z$ correspond to $\theta_{12}$, $\theta_{23}$ and 
$\theta_{13}$, respectively, where $\theta_{ij}$ are 
defined in the particle data group[11]. 
Since the Majorana CP-violation phases are irrelevant to the 
neutrino oscillations (flavor oscillations)[12], 
we neglected them. If we multiply the irrelevant phase matrix 
${\rm diag}(1,1,e^{-i\delta})$ from the right-hand side of $U$, 
we obtain the standard form[11]. The relation between 
the flavor eigenstates, $|\nu_\alpha\rangle$ $(\alpha=e,\mu,\tau)$, 
and the mass eigenstates, $|\nu_i\rangle$ $(i=1,2,3)$, is given by
\bea
|\nu_\alpha\rangle=U_{\alpha i}|\nu_i\rangle\;.
\ena

The evolution of the flavor eigenstates in matter with energy $E$ 
is given by
\bea
i\frac{d}{dx}|\nu_\beta(x)\rangle=
H(x)_{\beta \alpha}|\nu_\alpha(x)\rangle\;,
\ena
where Hamiltonian $H(x)_{\beta \alpha}$ is given by 
\bea
H(x)_{\beta \alpha}
=\frac{1}{2E}\left\{U_{\beta i}
\pmatrix{0&&\cr &\Delta m_{21}^2 &\cr 
&&\Delta m_{31}^2}_{ii}U_{i\alpha}^\dagger+
\pmatrix{a(x)&&\cr &0& \cr &&0}_{\beta\alpha}\right\}\;.
\ena
Here $\Delta m_{ij}^2 \equiv m_i^2-m_j^2$ with $m_i$ being 
the mass of $|\nu_i\rangle$, $G_F$ is the
Fermi coupling constant and  
\bea
a(x)\equiv 2\sqrt{2}G_F n_e(x) E=7.56\times 10^{-5}
\left(\frac{\rho(x)}{\rm g/cm^3}\right)
\left(\frac{Y_e}{0.5}\right)
\left(\frac{E}{\rm GeV}\right){\rm eV^2}\;,
\ena
where $n_e(x)$, $Y_e$ and $\rho(x)$ are 
the electron number density, the electron fraction 
and the matter density, respectively. For the electron 
fraction, we use $Y_e=0.5$.

We separate the matter density fluctuation from its average $\bar a$, 
\bea
\delta a(x) \equiv a(x)-\bar a\;,
\ena
and consider the deviation $\delta a(x)$ as a perturbative 
term. That is, we solve the evolution equation by treating 
$\delta a(x)L/2E$ and $\Delta m_{21}^2L/2E$ as perturbative 
terms, because they are small for most of the cases of 
planned neutrino factories. 
The validity of this perturbation was discussed by Ota and Sato[9]. 
They in fact showed that the transition probability of the neutrino 
oscillation is well reproduced if the 1st order perturbation 
with respect to the symmetric matter profile, $\delta a(x)_s L/2E$,
is taken into account, where it is assumed that 
the symmetric matter profile is well 
approximated by the preliminary reference earth model (PREM)[10], 
for $L=3000$km, $L=7332$km and $L=12000$km.

\vskip 2mm
\noindent
(a) The definition of Hamiltonian 

Following the work by Ota and Sato[9], we 
divide $H(x)$ into the unperturbed part $H_{00}$ and 
perturbed parts, $H_{01}$ and $H_1$
\bea
H=H_{00}+H_{01}+H_{1}(x)\;,
\ena
where
\bea
H_{00}&\equiv&\frac{1}{2E}e^{i\theta_{y} \lambda_7}
{\rm diag }(1,1,e^{i\delta})\nonumber\\
&&\times \pmatrix{
\Delta m_{31}^2 s_{z}^2+\bar a+\Delta 
m_{21}^2 s_{x}^2 c_{z}^2 & 0 & 
\Delta m_{31}^2 s_{z} c_{z}-\Delta m_{21}^2 s_{x}^2 s_{z} c_{z}\cr
0&\Delta m_{21}^2 c_{x}^2 &0\cr 
\Delta m_{31}^2 s_{z} c_{z}-\Delta m_{21}^2 s_{x}^2 s_{z} c_{z} &
0& \Delta m_{31}^2 c_{z}^2+\Delta m_{21}^2 s_{x}^2 s_{z}^2}\nonumber\\
&&\times {\rm diag }(1,1,e^{-i\delta}) 
e^{-i\theta_y \lambda_7}\;,\nonumber\\
H_{01}&\equiv& \frac{\Delta m_{21}^2}{2E} 
s_{x} c_{x} e^{i\theta_{y} \lambda_7}
\pmatrix{1&0&0\cr 0&1&0\cr 0&0&e^{i\delta}} 
\pmatrix{0&c_{z}&0\cr c_{z}&0&-s_z\cr 0&-s_{z}&0}
\pmatrix{1&0&0\cr 0&1&0\cr 0&0&e^{-i\delta}}
e^{-i\theta_{y} \lambda_7}\;,\nonumber\\
H_1(x)&\equiv&\frac{\delta a(x)}{2E}\pmatrix{1&&\cr &0& \cr &&0}\;.
\ena

For the later calculations, we use the following quantities;
\bea
\tilde U_0&=&e^{i\theta_{y} \lambda_7}
{\rm diag(1,1,e^{i\delta})} e^{i{ \theta_{\tilde z}} \lambda_5}
=\pmatrix{c_\tp& 0 & s_{\tilde {z}}\cr 
-s_{y} s_{\tilde {z}} e^{i\delta}& c_{y} & s_{y} c_{\tilde {z}}
 e^{i\delta}\cr
-c_{y} s_{\tilde {z}} 
e^{i\delta}& -s_{y} & c_{y} c_{\tilde {z}}
 e^{i\delta}\cr}\;,\nonumber\\
\tan 2 \theta_{\tilde z}&=&\frac{s_{2z}
 (\Delta m_{31}^2-\Delta m_{21}^2 s_{x}^2)}
 {c_{2z}(\Delta m_{31}^2-\Delta m_{21}^2 s_{x}^2)- \bar a}
\;,\nonumber\\
\lambda_\pm&=&\frac{1}{2}\left( 
\Delta m_{31}^2+\Delta m_{21}^2 s_{x}^2+\bar a 
\right.\;
\nonumber\\
&&\hskip 1mm \left. \pm \sqrt{
\{c_{2 z}(\Delta m_{31}^2-\Delta m_{21}^2 s_{x}^2)-\bar a\}^2
+s_{2z}^2(\Delta m_{31}^2-\Delta m_{21}^2 s_{x}^2)^2}
\right)\;.
\ena
We also define
\bea
a_\pm&=&\frac{\lambda_\pm}{2E}\;,
\;\;\;a_0=\frac{\Delta m_{21}^2c_{x}^2}{2E}\;.
\ena
With $a_i$ ($i=\pm,0$), we define
\bea
k_1&=&a_0 -a_-\;,\;\;k_2=a_+-a_0\;,\nonumber\\
k&=&k_2+k_1=a_+ -a_-\;,\;\;
\nonumber\\
\phi_{1\pm}&=&e^{-ia_0 L}\pm e^{-ia_- L}\;,
\;\; \phi_{2\pm}=e^{-ia_+ L}\pm e^{-ia_0 L}\;,
\nonumber\\
\phi_{\pm}&=& e^{-ia_+ L}\pm e^{-ia_- L}\;,
\ena
and 
\bea
P(x)=\pmatrix{e^{-ia_-x}&0&0\cr 0&e^{-ia_0x}&0\cr
   0&0&e^{-ia_+x}\cr} \;.
\ena

\vskip 2mm
\noindent
(b) Interaction representation

In the interaction representation, the interaction Hamiltonians 
are given by 
\bea
H_{01}(x)_I\equiv e^{iH_{00}x}H_{01}(x)e^{-iH_{00}x}\;,\;\;
H_{1}(x)_I\equiv e^{iH_{00}x}H_{1}(x)e^{-iH_{00}x}\;,
\ena
and the wave functions are presented by 
$|\nu_\alpha(x)\rangle_I\equiv 
(e^{iH_{00}x})_{\alpha\beta}|\nu_\beta(x)\rangle$ 
with $|\nu_\alpha(0)\rangle_I=|\nu_\alpha(0)\rangle$. 
The evolution equation of $|\nu_\alpha(x)\rangle_I$ becomes
\bea
i\frac{d}{dx}|\nu_\alpha(x)\rangle_I
 =(H_{01}(x)_I+H_{1}(x)_I)_{\alpha\beta}|
\nu_\beta(x)\rangle_I\;.
\ena
The solution is given by
\bea
|\nu_\alpha(L)\rangle_I
= T\left({\rm exp}\left[(-i)\int_0^L dx H_{01}(x)_I+H_{1}(x)_I \right]
\right)_{\alpha\beta}|\nu_\beta(0)\rangle\;,
\ena
where $ T$ means the time ordered product.
Then 
\bea
|\nu_\alpha(L)\rangle&=&\left[e^{-iH_{00}L}\right]_{\alpha\beta}
|\nu_\beta(L)
\rangle_I
\nonumber\\
&=&\left[e^{-iH_{00}L}\right]_{\alpha\gamma}\Biggl[1
+(-i)\int_0^L dx H_{01}(x)_I
+(-i)\int_0^L dx H_1(x)_I\nonumber\\
&&+(-i)^2\int_0^L dx\int_0^x dy H_{01}(x)_IH_{01}(y)_I
+(-i)^2\int_0^L dx\int_0^x dy H_1(x)_IH_1(y)_I\nonumber\\
&&+(-i)^2\int_0^L dx\int_0^x dy \left( H_{01}(x)_IH_1(y)_I
+ H_1(x)_IH_{01}(y)_I\right)
+\cdots\Biggr]_{\gamma\beta}|\nu_\beta(0)\rangle\nonumber\\
&\equiv&\left(
S_{00}+S_{01}+S_1^{(1)}+S_{01,01}+S_{1}^{(2)}+S_{01,1}+\cdots
\right)_{\alpha\beta}|\nu_\beta(0)\rangle\;.
\ena

The transition probability from one flavor eigenstate $\alpha$ 
to another $\beta$ is given by 
\bea
P(\nu_\alpha \to \nu_\beta)&=&|S(L)_{\alpha\beta}|^2\nonumber\\
&=& |(S_{00})_{\alpha\beta}|^2\nonumber\\ 
&+&2{\rm Re}\left[(S_{00})_{\alpha\beta}(S_{01})_{\alpha\beta}^\ast 
+(S_{00})_{\alpha\beta}(S_{1}^{(1)})_{\alpha\beta}^\ast\right]
\nonumber\\
&+&|(S_{01})_{\alpha\beta}|^2 +|(S_{1}^{(1)})_{\alpha\beta}|^2 
+2{\rm
Re}\left[(S_{01})_{\alpha\beta}(S_{1}^{(1)})_{\alpha\beta}^\ast\right]
\nonumber\\
&+&2{\rm Re}\left[(S_{00})_{\alpha\beta}(S_{01,01})_{\alpha\beta}^\ast
+(S_{00})_{\alpha\beta}(S_{1}^{(2)})_{\alpha\beta}^\ast
+(S_{00})_{\alpha\beta}(S_{01,1})_{\alpha\beta}^\ast\right]
\nonumber\\
&+&\cdots\;.
\ena
T-violation is defined by
\bea
\Delta P_{\nu_\alpha \nu_\beta}^{T}
=P(\nu_\alpha \to \nu_\beta)-P(\nu_\beta \to \nu_\alpha)\;,
\ena
and is evaluated with use of the probabilities defined in Eq.(17).

\section{The 0th and 1st order contribution from $H_{01}(x)$}

Ota and Sato[9] calculated the 1st order contribution from 
$H_{01}(x)$. Here, we give a brief derivation of their results 
which are needed to discuss the higher order calculation.

\noindent
(a) The S-matrix (0th and 1st order)

The unperturbed Hamiltonian $H_{00}$ is diagonalized explicitly 
and is given by 
\bea
H_{00}\equiv \tilde U_0
{\rm diag}(a_-,a_0,a_+)
\tilde U_0^\dagger\;,
\ena
where $a_\pm$ and $a_0$ are defined in Eq.(10). 

The S-matrix for $H_{00}$ is easily obtained as 
\bea
&&S_{00}=e^{-iH_{00}L}=\tilde{U}_0 P(L)
\tilde{U}_0^\dagger\nonumber\\
&=&\frac12
\pmatrix{\phi_+-c_{2\tp}\phi_-
& s_{y} s_{2\tp} e^{-i\delta}\phi_- 
& c_{y} s_{2\tp} e^{-i\delta}\phi_- \cr
  s_{y} s_{2\tp} e^{i\delta}\phi_- 
& \phi_++s_y^2c_{2\tp}\phi_--c_y^2(\phi_{2-}-\phi_{1-})
& s_{2y}
\left(c_{\tp}^2 \phi_{2-}-s_{\tp}^2 \phi_{1-}\right)\cr
  c_{y} s_{2\tp} e^{i\delta}\phi_- 
& s_{2y}\left(c_{\tp}^2 \phi_{2-}-s_{\tp}^2 \phi_{1-}
\right)
& \phi_++c_y^2c_{2\tp}\phi_--s_y^2(\phi_{2-}-\phi_{1-})\cr
}\;,
\nonumber\\
\ena
where $\phi_\pm$ and $\phi_{i\pm}$ ($i=1,2$) are defined in Eq.(11). 
The 1st order term of $H_{01}$ defined in Eq.(8) is 
given by 
\bea
S_{01}&=&\tilde{U}_0 
\pmatrix{0&{\cal A}_{01}&0\cr {\cal A}_{01}&0&
{\cal B}_{01}\cr 0&{\cal B}_{01}&0}
\tilde{U}_0^\dagger\nonumber\\
&=&\pmatrix{
0 & c_y{\cal P}_{01} 
& -s_y {\cal P}_{01}\cr 
c_y {\cal P}_{01}
& -s_{2y} c_\delta {\cal Q}_{01}
&(e^{i\delta}s_y^2-e^{-i\delta}c_y^2)
 {\cal Q}_{01}\cr
-s_y {\cal P}_{01}
&(e^{-i\delta}s_y^2-e^{i\delta}c_y^2) {\cal Q}_{01}
&s_{2y} c_\delta{\cal Q}_{01}\cr}\;,
\ena
where
\bea
{\cal P}_{01}&=&c_{\tp}{\cal A}_{01}+s_{\tp}{\cal B}_{01}
\;,\;{\cal Q}_{01}=s_{\tp}{\cal A}_{01}-c_{\tp}{\cal B}_{01}\;,
\nonumber\\
{\cal A}_{01}&=& \frac{\Delta m^2_{21}}{4E}\frac{s_{2x}c_{z-\tp}}
{k_1}\phi_{1-}\;,\quad 
{\cal B}_{01}= -\frac{\Delta m^2_{21}}{4E}\frac{s_{2x}s_{z-\tp}}
{k_2}\phi_{2-}\;.
\ena

\vskip 2mm
\noindent
(b) The oscillation probabilities (0th and 1st order)

The oscillation probability in the 0th order 
$P^{(00)}(\nu_\alpha \to \nu_\beta)$ is 
given by $|(S_{00})_{\alpha\beta}|^2$,
\bea
P^{(00)}(\nu_e \to \nu_\mu)&=&P^{(00)}(\nu_\mu \to \nu_e)
=s_y^2 s_{2\tp}^2 \sin^2{\frac{kL}{2}}\;,\nonumber\\
P^{(00)}(\nu_e \to \nu_\tau)&=&P^{(00)}(\nu_\tau \to \nu_e)
=c_y^2 s_{2\tp}^2 \sin^2{\frac{kL}{2}}\;,
\nonumber\\
P^{(00)}(\nu_\mu \to \nu_\tau)&=&
P^{(00)}(\nu_\tau \to \nu_\mu)\nonumber\\
&=&s_{2y}^2\left\{
        s_{\tp}^2\sin^2{\frac{k_1L}{2}}
        +c_{\tp}^2\sin^2{\frac{k_2L}{2}}
        -s_{\tp}^2c_{\tp}^2\sin^2{\frac{kL}{2}}\right\}
        \;,
\ena
with $k$, $k_1$ and $k_2$ defined in Eq.(11). 

The probability for antineutrinos 
$P^{(00)}(\bar {\nu_\alpha} \to \bar{\nu_\beta})$ is obtained 
by taking $\delta \to -\delta$ and $\bar a \to -\bar a$. 

For the higher order terms, we consider only the contribution to 
T-violation.  
The 1st order term from $H_{01}$, i.e., the $\Delta m_{21}^2L/2E$ 
term is solely by the CP violation phase $\delta$ and is given 
by
\bea
\left(\Delta P^{T}_{\nu_e \nu_\mu}\right)_{s_\delta}
=-\frac{\Delta m^2_{21}}{E}s_{2x}s_{2y}s_{2\tp}s_\delta 
\left[\frac{c_{\tp}c_{z-\tp}}{k_1}+\frac{s_{\tp}s_{z-\tp}}{k_2}
\right]
\sin \frac{k_1L}2 \sin \frac{k_2L}2 \sin \frac{kL}2 \;.\,
\ena
where we used $\sin k_1L+\sin k_2L -\sin kL=
4\sin (k_1L/2)\sin (k_2L/2)\sin (kL/2)$, which is proved by using 
$k=k_1+k_2$. 
This formula includes the constant matter effect.

Similarly, we find
\bea
\left(\Delta P^{T}_{\nu_\mu  \nu_\tau}\right)_{s_\delta}
=\left( \Delta P^{T}_{\nu_\tau\nu_e}\right)_{s_\delta}
=\left(\Delta P^{T}_{\nu_e \nu_\mu}\right)_{s_\delta}\;,
\ena
which is valid for the constant media as stated in the paper  
by Krastev and Petcov[5]. 

\section{The contribution from the matter term, $H_1(x)$}

In this section, we give the general formula to evaluate 
the n-th order effects of matter, i.e.,  the  
$(\delta a(x)L/2E)^n$ order. Then, we evaluate the contributions of
up to the 3rd order and discuss the general properties of the 
effects.

\vskip 2mm
\noindent
(a) S-matrix elements

The interaction Hamiltonian with matter in the 
interaction representation is expressed 
by
\bea
H_1(x)_I=\frac{\delta a(x)}{2E}\tilde{U}_0
\pmatrix{c_{\tp}^2&0&s_{\tp}c_{\tp}e^{-ikx}\cr
          0&0&0\cr s_{\tp}c_{\tp}e^{ikx}&0&s_{\tp}^2\cr}
\tilde{U}_0^\dagger\;.
\ena
Then, the n-th order matter perturbation is given by
\bea
S_1^{(n)}&=&e^{-iH_{00}L}(-i)^n \int_0^L dx_1 
\cdots \int_0^{x_{n-1}} dx_n  H_1(x_1)_IH_{1}(x_2)_I
\cdots H_{1}(x_n)_I\nonumber\\
& =&\tilde{U}_0(-i)^n P(L)\left\{\int_0^L dx_1 
\cdots \int_0^{x_{n-1}} dx_n  \frac{\delta a(x_1)}{2E}
\cdots \frac{\delta a(x_n)}{2E}\right. \nonumber\\
&&\times \left.
\pmatrix{c_{\tp}^2&0&s_{\tp}c_{\tp}e^{-ikx_1}\cr
          0&0&0\cr s_{\tp}c_{\tp}e^{ikx_1}&0&s_{\tp}^2\cr}
          \cdots
\pmatrix{c_{\tp}^2&0&s_{\tp}c_{\tp}e^{-ikx_n}\cr
          0&0&0\cr s_{\tp}c_{\tp}e^{ikx_n}&0&s_{\tp}^2\cr}
          \right\} \tilde{U}_0^\dagger\;.
\ena
Therefore, the S-matrix is written in general by
\bea
S_1^{(n)}=\tilde{U}_0\pmatrix{{\cal E}_1^{(n)}&0&{\cal C}_1^{(n)}\cr
0&0&0\cr {\cal D}_1^{(n)}&0&{\cal F}_1^{(n)}}\tilde{U}_0^\dagger
=\frac12
\pmatrix{\alpha_{1,n}^{(+)}&e^{-i\delta}s_y\beta_{1,n}^{(+)}&
e^{-i\delta}c_y \beta_{1,n}^{(+)}\cr
e^{i\delta}s_y\beta_{1,n}^{(-)}&s_y^2\alpha_{1,n}^{(-)}
&s_yc_y\alpha_{1,n}^{(-)}\cr
e^{i\delta}c_y \beta_{1,n}^{(-)}&s_yc_y\alpha_{1,n}^{(-)}
&c_y^2\alpha_{1,n}^{(-)}\cr}\;,
\ena
where
\bea
\alpha_{1,n}^{(\pm)}&=&{\cal E}_{1}^{(n)}+{\cal F}_{1}^{(n)}
\pm \left(c_{2\tp}({\cal E}_{1}^{(n)}-{\cal F}_{1}^{(n)})+
s_{2\tp}({\cal C}_{1}^{(n)}+{\cal D}_{1}^{(n)})\right)  \;,
\nonumber\\
\beta_{1,n}^{(\pm)}&=&-s_{2\tp}({\cal E}_{1}^{(n)}-{\cal F}_{1}^{(n)})
+c_{2\tp}({\cal C}_{1}^{(n)}+{\cal D}_{1}^{(n)})\pm 
({\cal C}_{1}^{(n)}-{\cal D}_{1}^{(n)}) \;.
\ena

Below, we show their explicit forms, which are 
needed to evaluate the matter effect to T-violation  
up to the 3rd order. The 1st order terms are 
\bea
{\cal C}_{1}^{(1)}&=&(-i)e^{-ia_-L}\int_0^L dx
  \frac{\delta a(x)}{2E} s_\tp c_\tp e^{-ikx}\;,
  \nonumber\\
{\cal D}_{1}^{(1)}&=&(-i)e^{-ia_+L}\int_0^L dx 
  \frac{\delta a(x)}{2E} s_\tp c_\tp e^{ikx}\;,
  \nonumber\\
{\cal E}_{1}^{(1)}&=&(-i)e^{-ia_-L}\int_0^L dx 
  \frac{\delta a(x)}{2E}c_\tp^2 \;,
  \nonumber\\
{\cal F}_{1}^{(1)}&=&(-i)e^{-ia_+L}\int_0^L dx 
  \frac{\delta a(x)}{2E}s_\tp^2 \;.
\ena
The 2nd order terms are
\bea
{\cal C}_{1}^{(2)}&=&(-i)^2e^{-ia_-L}\int_0^L dx\int_0^{x} dy
  \frac{\delta a(x)}{2E}\frac{\delta a(y)}{2E}
  s_\tp c_\tp\left(s_\tp^2 e^{-ikx}+c_\tp^2 e^{-iky}\right)\;,
  \nonumber\\
{\cal D}_{1}^{(2)}&=&(-i)^2e^{-ia_+L}\int_0^L dx\int_0^{x} dy
  \frac{\delta a(x)}{2E}\frac{\delta a(y)}{2E}
  s_\tp c_\tp \left(c_\tp^2 e^{ikx}+s_\tp^2 e^{iky}\right)\;,
  \nonumber\\
{\cal E}_{1}^{(2)}&=&(-i)^2e^{-ia_-L}\int_0^L dx\int_0^{x} dy
  \frac{\delta a(x)}{2E}\frac{\delta a(y)}{2E}
  c_\tp^2\left(c_\tp^2 +s_\tp^2 e^{-ik(x-y)}\right)\;,
 \nonumber\\
{\cal F}_{1}^{(2)}&=&(-i)^2e^{-ia_+L}\int_0^L dx\int_0^{x} dy
  \frac{\delta a(x)}{2E}\frac{\delta a(y)}{2E}
   s_\tp^2\left(s_\tp^2 +c_\tp^2 e^{ik(x-y)}\right )\;.
\ena
Finally the 3rd order terms are
\bea
{\cal C}_{1}^{(3)}&=&(-i)^3e^{-ia_-L}\int_0^L dx\int_0^{x} dy
 \int_0^{y} dz
 \frac{\delta a(x)}{2E}\frac{\delta a(y)}{2E}\frac{\delta a(z)}{2E}
 \nonumber\\
&& \times s_\tp c_\tp \left(s_\tp^4 e^{-ikx}+s_\tp^2c_\tp^2( e^{-iky}+
 e^{-ik(x-y+z)})+c_\tp^4 e^{-ikz}\right)\;,
\nonumber\\
{\cal D}_{1}^{(3)}&=&(-i)^3e^{-ia_+L}\int_0^L dx\int_0^{x} dy
 \int_0^{y} dz
 \frac{\delta a(x)}{2E}\frac{\delta a(y)}{2E}\frac{\delta a(z)}{2E}
 \nonumber\\
 &&\times s_\tp c_\tp\left( c_\tp^4 e^{ikx}+s_\tp^2c_\tp^2( e^{iky}+
 e^{ik(x-y+z)})+s_\tp^4 e^{ikz}\right)\;.
\ena
It should be noted that ${\cal D}_{1}^{(n)}$ and 
${\cal F}_{1}^{(n)}$ are derived from ${\cal C}_{1}^{(n)}$ and 
${\cal E}_{1}^{(n)}$, by exchanging between $s_\tp$ and $c_\tp$, and 
also $a_-$ and $a_+$. 

In order to examine the general properties of these quantities, 
we divide the matter fluctuation $\delta a(x)$ into the 
symmetric part, $\delta a(x)_s$, and the asymmetric 
part, $\delta a(x)_a$, and then expand them 
in terms of Fourier cosine series,
\bea
\delta a(x)_s&=&\sum_{n\neq 0}a_{2n} e^{-iq_{2n} x}\;,
\nonumber\\
\delta a(x)_a &=&\sum_{n} a_{2n+1} e^{-iq_{2n+1} x}\;,
\;\;q_n=\frac{\pi n}{L}\;,
\ena 
where $a_n$ is a real number satisfying $a_{-n}=a_n$. 
Here, we excluded $n=0$ in the expression of $\delta a(x)_s$,
because 
$\delta a(x)$ is the deviation from the average of $a(x)$. 
Below, we consider the both profiles together. 

The 1st order terms are
\bea
{\cal C}_{1}^{(1)}+{\cal D}_{1}^{(1)}
&=&\frac{s_{2\tp}}{2E}\left(
  \sum_{n\neq 0}\frac{a_{2n}}{k+q_{2n}}\right)\phi_-\;,
  \nonumber\\
{\cal C}_{1}^{(1)}-{\cal D}_{1}^{(1)}&=&
-\frac{s_{2\tp}}{2E}\left(
\sum_{n}\frac{a_{2n+1}}{k+q_{2n+1}}\right)\phi_+\;,
\nonumber\\
{\cal E}_{1}^{(1)}&=&{\cal F}_{1}^{(1)}=0\;.
\ena
The 2nd order terms are 
\bea
{\cal C}_{1}^{(2)}+{\cal D}_{1}^{(2)}&=&
\frac{s_{2\tp}c_{2\tp}}{(2E)^2}\left(
\sum_{m+n={\rm even}}\frac{a_na_m}{(k+q_n)(k+q_n+q_m)}\right)
\phi_-\;,
\nonumber\\
{\cal C}_{1}^{(2)}-{\cal D}_{1}^{(2)}&=&
-\frac{s_{2\tp}c_{2\tp}}{(2E)^2}\left(
\sum_{m+n={\rm odd}}\frac{a_na_m}{(k+q_n)(k+q_n+q_m)}\right)\phi_+
\;,
\nonumber\\
{\cal E}_{1}^{(2)}-{\cal F}_{1}^{(2)}&=&
\frac{s_{2\tp}^2}{8(2E)^2}\left\{ \left(
\sum_{n,m}
 \frac{((-1)^{n}+1)((-1)^m+1)a_n a_m }{(k+q_n)(k+q_m)}
\right)\phi_- \right.\nonumber\\
&&\hskip 1.5cm +\left. 2L\left(\sum_{n}\frac{a_na_{-n}}{k+q_n}
\right)i\phi_+
\right\}\;.
\ena
Finally, the 3rd order terms are 
\bea
&&{\cal C}_{1}^{(3)}-{\cal D}_{1}^{(3)}=
\frac{s_{2\tp}}{2(2E)^3}\left\{ \left[ 
c_{2\tp}^2\sum_{n,m,l}
\frac{((-1)^{n+m+l}-1)a_na_ma_l}
{(k+q_n)(k+q_n+q_m)(k+q_n+q_m+q_l)}\right.\right.
\nonumber\\
&&\hskip 2mm
-\frac{s_{2\tp}^2}{8}\left(
\sum_{n,m,l}
\frac{((-1)^{n+m+l}-1)a_na_ma_l}
{(k+q_n)(k+q_m)(k+q_l)}
+\sum_{l}\frac{((-1)^{l}-1)a_l}{(k+q_l)}
\sum_{m\neq -n}\frac{a_na_{m}}{(k+q_n)(k+q_m)}\right.
\nonumber\\
&&\hskip 4mm \left.\left.\left. +2\sum_{n,m,l}
\frac{((-1)^{l}-1)a_l}{(k+q_l)^2}
\sum_n\frac{a_na_{-n}}{(k+q_n)}\right)\right]\phi_+
-\frac{s_{2\tp}^2}{4}L
\sum_{l}\left[\frac{((-1)^{l}-1)a_l}{(k+q_l)}
\sum_n\frac{La_na_{-n}}{(k+q_n)}\right]i\phi_-\right\}\;.
\nonumber\\
\ena

\vskip 2mm
\noindent
(b) The general properties

We first discuss the n-th order contributions of the matter 
to the transition probabilities, which are obtained by 
computing 
\bea
P^{(1,n)}(\nu_\alpha \to \nu_\beta)=
{\rm Re}\left(2(S_{00})_{\alpha\beta}(S_1^{(n)*})_{\alpha\beta}
+\sum_{l+m=n}(S_{1}^{(l)})_{\alpha\beta}(S_1^{(m)*})_{\alpha\beta}
\right)\;.
\ena
Since the $(S_1^{(n)})_{\mu\tau}=(S_1^{(n)})_{\tau\mu}$, 
we conclude that 
\bea
P^{(1,n)}(\nu_\mu \to\nu_\tau)=P^{(1,n)}(\nu_\tau \to\nu_\mu)\;.
\ena
Therefore, the matter itself does not give any effect to 
T-violation  for $\nu_\mu \to\nu_\tau$  channel.

For other channels,  
\bea
P^{(1,n)}(\nu_e \to\nu_\mu)=\frac{s_y^2}{4}{\rm Re}
\left(2 s_{2\tp}\phi_-\beta_{1,n}^{(+)*}+\sum_{l+m=n}
\beta_{1,l}^{(+)}\beta_{1,m}^{(+)*}
\right)\;,
\ena
and $P^{(1,n)}(\nu_\mu \to \nu_e)$ is obtained by changing the 
superscript $(+)$ to $(-)$. The transition probabilities for 
$\nu_e \to \nu_\tau$ and $\nu_\tau \to \nu_e$ are obtained 
from those for $\nu_e \to \nu_\mu$ and 
$\nu_\mu \to \nu_e$, by changing the coefficient $s_y^2$ with 
$c_y^2$. 

We consider the 1st order of the asymmetric matter fluctuation, 
which contributes only ${\cal C}_{1}^{(1)}-{\cal D}_{1}^{(1)}$,  
in contrast to the ${\cal C}_{1}^{(1)}+{\cal D}_{1}^{(1)}$ 
which is due to the symmetric matter profile. In addition, 
${\cal E}_{1}^{(1)}=0$ and ${\cal F}_{1}^{(1)}=0$. Therefore, 
the asymmetric matter fluctuation gives  $\alpha_{1,1}^{(-)}=0$, 
which shows that it does not contribute 
to $\nu_\mu \to \nu_\tau$ channel. Also we find 
$\beta_{1,1}^{(\pm)}=\pm({\cal C}_{1}^{(1)}-{\cal D}_{1}^{(1)})$, 
which is proportional to $\phi_+$. Since 
\bea
\phi_-\phi_+^*=-2i\sin kL\;,
\ena
which implies the vanishing contribution. 
Thus we conclude that the asymmetric matter profile does not 
contribute to the transition probability in the 1st order. 
It contributes to the transition probability in the 2nd order.

\vskip 2mm
\noindent
(c) T-violation   only from the matter 

\vskip 2mm
\noindent 
(c-1) The 1st order effect

The 1st order contribution to $\Delta P^T_{\nu_e \nu_\mu}$ is from 
the interference between $S_{00}$ and $S_1^{(1)}$ as we 
see in Eq.(17). 
In this case, the ${\cal C}_{1}^{(1)}-{\cal D}_{1}^{(1)}$ 
in $S_1^{(1)}$ 
contributes to T-violation. Since 
${\cal C}_{1}^{(1)}-{\cal D}_{1}^{(1)}$ is proportional 
to $\phi_+$, while $S_{00}$ to $\phi_-$, 
T-violation  is proportional to ${\rm
Re}[\phi_-\phi_+^*]$, 
which is zero.
Therefore, there is no 1st order effect to T-violation. 

\vskip 2mm
\noindent 
(c-2) The 2nd order effect

The 2nd order effect comes from $|S_1^{(1)}|^2$ and 
the interference between $S_{00}$ and $S_1^{(2)}$. 
Similarly to the 1st order contribution, 
Re[$\phi_-({\cal C}_{1}^{(2)*}-{\cal D}_{1}^{(2)*})]$ in 
$S_1^{(2)}$ contributes to the asymmetry. Since 
${\cal C}_{1}^{(2)}-{\cal D}_{1}^{(2)}$ is proportional 
to $\phi_+$, this term does not contribute. As for 
$|S_1^{(1)}|^2$ term, Re$[({\cal E}_{1}^{(1)}-{\cal F}_{1}^{(1)})
({\cal C}_{1}^{(1)*}-{\cal D}_{1}^{(1)*})]$ and also 
Re$[({\cal C}_{1}^{(1)}+{\cal D}_{1}^{(1)}) 
({\cal C}_{1}^{(1)*}-{\cal D}_{1}^{(1)*})]$ contribute. 
Since ${\cal C}_{1}^{(1)}+{\cal D}_{1}^{(1)}$ is proportional to 
$\phi_-$ and ${\cal E}_{1}^{(1)}-{\cal F}_{1}^{(1)}$ consists 
of $\phi_-$ and $i\phi_+$, they do not contribute.
As a result, there is no 2nd order effect.

\vskip 2mm
\noindent 
(c-3) The 3rd order effect

The 3rd order effect comes from $S_{00}S_1^{(3)*}$ and 
$S_1^{(1)}S_1^{(2)*}$. Since 
${\cal C}_{1}^{(3)}-{\cal D}_{1}^{(3)}$ contains $\phi_+$ and 
$i\phi_-$, no contribution arises from $S_{00}S_1^{(3)*}$ 
term. Similarly, for $S_1^{(1)}S_1^{(2)*}$ terms, the 
terms contributing to T-violation   are 
Re$[({\cal E}_{1}^{(1)}-{\cal F}_{1}^{(1)})
({\cal C}_{1}^{(2)*}-{\cal D}_{1}^{(2)*})]$, 
Re$[({\cal C}_{1}^{(1)}+{\cal D}_{1}^{(1)}) 
({\cal C}_{1}^{(2)*}-{\cal D}_{1}^{(2)*})]$ 
and terms that the superscript 1 and 2 are exchanged. 
Again it is clear that these terms do not contribute. 

We conclude that there is no contribution to T-violation  
through the matter terms 
up to the 3rd order, i.e., $(\delta a(x)L/2E)^3$ terms. 
The vanishing of the 3rd order term is sufficient enough 
to deal with the actual situation. Thus we do not pursue the 
further investigation, although we expect that the higher terms 
will not contribute either.

\section{The second order contribution from $H_{01}$} 

We consider the $(\Delta m_{21}^2L/2E)^2$ effect. 
For this, there are two contributions. One is from 
$|(S_{01})_{\beta\alpha}|^2$ and the other is 
from $2{\rm Re}[(S_{00})_{\beta\alpha}(S_{01,01})_{\beta\alpha}^*]$. 

We first compute $S_{01,01}$ defined in Eq.(17)
\bea
S_{01,01}
=\tilde{U}_0 P(L)(-i)^2 \int_0^L dx \int_0^x dy
\left[P(-x) \tilde{U}_0^\dagger H_{01} \tilde{U}_0 
P(x-y) \tilde{U}_0^\dagger H_{01} \tilde{U}_0 P(y)
\right]\tilde{U}_0^\dagger\;.
\ena
The quantity of the bracket becomes 
$(\Delta m^2_{21} s_{2x}/4E)^2$ multiplied by 
\bea
\pmatrix{
c_{z-\tp}^2 e^{-ik_1(x-y)} & 0& 
-\frac{1}{2}s_{2(z-\tp)}e^{-i(k_1x+k_2y)}\cr
0&c_{z-\tp}^2e^{ik_1(x-y)}+s_{z-\tp}^2 e^{-ik_2(x-y)}&
0\cr
-\frac{1}{2}s_{2(z-\tp)}e^{i(k_2x+k_1y)}&0
&s_{z-\tp}^2 e^{ik_2(x-y)} \cr}\;.
\ena
Then, we find
\bea
S_{01,01}&=&\tilde{U}_0\pmatrix{
{\cal E}_{01,01}& 0& {\cal C}_{01,01}\cr 
0&{\cal G}_{01,01}&0\cr
{\cal D}_{01,01}& 0& {\cal F}_{01,01}}
\tilde{U}_0^\dagger\nonumber\\
&=&\frac12
\pmatrix{\alpha_{01,01}^{(+)}&e^{-i\delta}s_y\beta_{01,01}^{(+)}&
e^{-i\delta}c_y \beta_{01,01}^{(+)}\cr
e^{i\delta}s_y\beta_{01,01}^{(-)}&s_y^2\alpha_{01,01}^{(-)}
+2c_y^2 {\cal G}_{01,01}&s_yc_y\alpha_{01,01}^{(-)}
-s_{2y}{\cal G}_{01,01}\cr
e^{i\delta}c_y \beta_{01,01}^{(-)}&s_yc_y\alpha_{01,01}^{(-)}
-s_{2y}{\cal G}_{01,01}&
c_y^2\alpha_{01,01}^{(-)}+2s_y^2{\cal G}_{01,01}\cr}\;,
\ena
where
\bea
\alpha_{01,01}^{(\pm)}&=&({\cal E}_{01,01}+{\cal F}_{01,01})
\pm \left(c_{2\tp}({\cal E}_{01,01}-{\cal F}_{01,01})+
s_{2\tp}({\cal C}_{01,01}+{\cal D}_{01,01})  \right)\;,
\nonumber\\
\beta_{01,01}^{(\pm)}&=&-s_{2\tp}({\cal E}_{01,01}-{\cal F}_{01,01})
+c_{2\tp}({\cal C}_{01,01}+{\cal D}_{01,01})\pm 
({\cal C}_{01,01}-{\cal D}_{01,01}) \;.
\ena

We compute the 2nd order $(\Delta m_{21}^2 L/2E)^2$ 
contribution to T-violation. Since
$(S_{01,01})_{\mu\tau}
=(S_{01,01})_{\tau\mu}$, there is no contribution to 
T-violation   for the $\nu_\mu$ and $\nu_\tau$ oscillation 
channels. For the $\nu_e$ and $\nu_\mu$ channels, the 
${\cal C}_{01,01}-{\cal D}_{01,01}$ contributes because 
\bea
2{\rm Re}[(S_{00})_{\mu e}(S_{01,01})_{\mu e}^\ast-
(S_{00})_{e\mu}(S_{01,01})_{e\mu}^\ast]
=s_y^2 s_{2\tp}{\rm Re}
[\phi_-({\cal C}_{01,01}^\ast-{\cal D}_{01,01}^\ast)]\;.
\ena
The same holds for the $\nu_e$ and $\nu_\tau$ channel. 

${\cal C}_{01,01}$ and ${\cal D}_{01,01}$ are given by 
\bea
{\cal C}_{01,01}&=& 
-\left(\frac{\Delta m^2_{21} s_{2x}}{4E}\right)^2
\frac{1}{2}s_{2(z-\tp)}
\left[\frac{1}{kk_2}\phi_--\frac{1}{k_1k_2}\phi_{1-}\right]\;,\nonumber\\
{\cal D}_{01,01}&=& -\left(\frac{\Delta m^2_{21} s_{2x}}{4E}\right)^2
\frac{1}{2}s_{2(z-\tp)}
\left[-\frac{1}{kk_1}\phi_-+\frac{1}{k_1k_2}
\phi_{2-}\right]\;.
\ena 
Since  
\bea
{\cal C}_{01,01}^\ast-{\cal D}_{01,01}^\ast
&=&
-\left(\frac{\Delta m^2_{21} s_{2x}}{4E}\right)^2
\frac{1}{2}s_{2(z-\tp)}\left[
\frac{1}{kk_2}\phi_-^\ast-\frac{1}{k_1k_2}\phi_{1-}^*
+\frac{1}{kk_1}\phi_-^\ast-\frac{1}{k_1k_2}\phi_{2-}^*
\right]\nonumber\\
&=&-\left(\frac{\Delta m^2_{21} s_{2x}}{4E}\right)^2
\frac{1}{2}s_{2(z-\tp)}
\left[\frac{1}{kk_2}-\frac{1}{k_1k_2}+\frac{1}{kk_1}
\right]\phi_-^\ast =0\;,
\ena
because $\phi_{2-}+\phi_{1-}=\phi_-$ and $k_2=k-k_1$, 
there is no contribution to these channels.

The $|(S_{01})_{\beta\alpha}|^2$ also give a null contribution 
because
\bea
|(S_{01})_{e \mu}|^2-|(S_{01})_{\mu e}|^2
&=&|c_y(c_{\tp}{\cal A}_{01}+s_{\tp}{\cal B}_{01})|^2-
|c_y(c_{\tp}{\cal A}_{01}+s_{\tp}{\cal B}_{01})|^2
=0\;,\nonumber\\
|(S_{01})_{e\tau}|^2-|(S_{01})_{\tau e}|^2
&=&|s_y(c_{\tp}{\cal A}_{01}+s_{\tp}{\cal B}_{01})|^2
-|s_y(c_{\tp}{\cal A}_{01}+s_{\tp}{\cal B}_{01})|^2=0
\;,\nonumber\\
|(S_{01})_{\mu \tau}|^2-|(S_{01})_{\tau\mu}|^2
&=&|(e^{i\delta}s_y^2-e^{-i\delta}c_y^2)
(s_{\tp}{\cal A}_{01}-c_{\tp}{\cal B}_{01})|^2\nonumber\\
&&-|(e^{-i\delta}s_y^2-e^{i\delta}c_y^2)
(s_{\tp}{\cal A}_{01}-c_{\tp}{\cal B}_{01})|^2
=0\;.
\ena

Therefore, we find no effect from the 2nd order term of $H_{01}$, 
\bea
\Delta P^{T(01,01)}_{\nu_e \nu_\mu}=
\Delta P^{T(01,01)}_{\nu_e \nu_\tau}
=\Delta P^{T(01,01)}_{\nu_\mu \nu_\tau} =0\;.
\ena

\section{The interference between $H_{01}$ and $H_1$}

In this section, we consider the 
$(\Delta m_{21}^2L/2E)(\delta a(x)L/2E)$ contribution to 
T-violation. 

The S-matrix is given by
\bea
S_{01,1}=\tilde U_0\pmatrix{0&{\cal A}_{01,1}&0\cr
   {\cal A'}_{01,1}&0&{\cal B'}_{01,1}\cr
   0&{\cal B}_{01,1}&0\cr}\tilde U_0^\dagger\;,
\ena
where
\bea
{\cal A}_{01,1}&=&(-i)^2\frac{\Delta m_{21}^2s_{2x}}{4E}
e^{-ia_-L}\int_0^L dx\int_0^x dy\frac{\delta a(x)}{2E}c_{\tp}
\left( c_{\tp}c_{z-\tp}e^{-ik_1y}-s_\tp s_{z-\tp}
e^{-i(kx-k_2y)}
\right)\;,
\nonumber\\
{\cal A'}_{01,1}&=&(-i)^2\frac{\Delta m_{21}^2s_{2x}}{4E}
e^{-ia_0L}\int_0^L dx\int_0^x dy\frac{\delta a(y)}{2E}c_{\tp}
\left( c_{\tp} c_{z-\tp}e^{ik_1x}-s_\tp  s_{z-\tp}
e^{i(ky-k_2x)}
\right)\;,
\nonumber\\
{\cal B}_{01,1}&=&(-i)^2\frac{\Delta m_{21}^2s_{2x}}{4E}
e^{-ia_+L}\int_0^L dx\int_0^x dy\frac{\delta a(x)}{2E}s_{\tp}
\left( -s_{\tp} s_{z-\tp}e^{ik_2y}+ c_\tp c_{z-\tp}
e^{i(kx-k_1y)}
\right)\;,
\nonumber\\
{\cal B'}_{01,1}&=&(-i)^2\frac{\Delta m_{21}^2s_{2x}}{4E}
e^{-ia_0L}\int_0^L dx\int_0^x dy\frac{\delta a(y)}{2E}s_{\tp}
\left( -s_{\tp} s_{z-\tp}e^{-ik_2x}+c_\tp c_{z-\tp}
e^{-i(ky-k_1x)}
\right)\;.
\nonumber\\
\ena
Then, the S-matrix is given with use of ${\cal A}_{01,1}$ and 
${\cal B}_{01,1}$ by
\bea
S_{01,1}=
\pmatrix{0&c_y{\cal P}_{01,1}&-s_y {\cal P}_{01,1}\cr
c_y {\cal P}_{01,1}'
&-s_yc_y(e^{i\delta} {\cal Q}_{01,1}
+e^{-i\delta} {\cal Q}_{01,1}') 
&e^{i\delta} s_y^2{\cal Q}_{01,1}
-e^{-i\delta} c_y^2{\cal Q}_{01,1}' \cr
-s_y {\cal P}_{01,1}'   
&-e^{i\delta} c_y^2{\cal Q}_{01,1}
+e^{-i\delta} s_y^2{\cal Q}_{01,1}' 
& s_yc_y(e^{i\delta} {\cal Q}_{01,1}
+e^{-i\delta} {\cal Q}_{01,1}')\cr}\;,
\ena
where
\bea
{\cal P}_{01,1}&=&(c_\tp {\cal A}_{01,1}+s_\tp {\cal B}_{01,1})\;,\; 
{\cal Q}_{01,1}=(s_\tp {\cal A}_{01,1}-c_\tp {\cal B}_{01,1})\;,
\nonumber\\
{\cal P}_{01,1}'&=&(c_\tp {\cal A}_{01,1}'+s_\tp {\cal
B}_{01,1}')\;,\; 
{\cal Q}_{01,1}'=(s_\tp {\cal A}_{01,1}'-c_\tp {\cal B}_{01,1}')\;.
\ena

\vskip 2mm
\noindent
(a) The contribution from the symmetric matter fluctuation

The interference terms between $\Delta m_{21}^2 L/2E$ and 
the symmetric matter fluctuation, $\delta a(x)_s$, contributes 
only for the term proportional to the CP phase, $\sin \delta$. 
There are two contributions. One is from the 
$S_{01}S_{1s}^{(1)*}$ and the other is $S_{00}S_{01,1s}^{*}$ 
terms.  We take the PREM[10] as a symmetric matter. Ota and Sato  
expanded the PREM in the cosine series[9], which is expressed as 
$\delta a(x)_s$ in Eq.(33) with appropriate values of coefficients. 
The $S_{1s}^{(1)*}$ is given by Eqs.(28) and (34), where we 
take only the even $n$ case, i.e.,  
from ${\cal C}_{1}^{(1)}+{\cal D}_{1}^{(1)}$. We find
\bea
&&2{\rm Re}[(S_{01})_{e \mu}(S_{1s})_{e \mu }^\ast
-(e \iff \mu)]
\nonumber\\
&&\hskip 5mm =-\frac{\Delta m^2_{21}s_{2x}s_{2y}s_{2\tp}s_\delta}{E}
\left(\frac{c_{z-\tp}c_{\tp}}{k_1}
+\frac{s_{z-\tp}s_{\tp}}{k_2}\right)c_{2\tp}
\nonumber\\
&&\hskip 10mm \times 
\left(\sum_{n=1,2,\cdots}\frac{a_{2n}k}{E(k^2-q_{2n}^2)}\right)
        \sin \frac{k_1L}2 
\sin \frac{k_2L}2 \sin \frac{kL}2 \;.
\ena

The term $S_{01,1s}$ is obtained once we compute 
${\cal A}_{01,1s}$ etc., which are given by 
\bea
{\cal A}_{01,1s}&=&{\cal A}_{01,1s}'=
\frac{\Delta m_{21}^2 s_{2x}}{4E} c_\tp \left\{
\left(\frac{c_\tp c_{z-\tp}}{k_1}+\frac{s_\tp s_{z-\tp}}{k_2}\right) 
\right. \nonumber\\
&&\times \left. 
\left(\sum_{n=1,2,\cdots}\frac{a_{2n}k_1}{E(k_1^2-q_{2n}^2)}\right)
\phi_{1-}
-\frac{s_\tp s_{z-\tp}}{k_2}\sum_{n=1,2,\cdots}
\left(\frac{a_{2n}k}{E(k^2-q_{2n}^2)}\right)\phi_{-}\right\}\;,
\nonumber\\
{\cal B}_{01,1s}&=&{\cal B}_{01,1s}'=
\frac{\Delta m_{21}^2s_{2x}}{4E}s_\tp\left\{
\left(\frac{c_\tp c_{z-\tp}}{k_1}+\frac{s_\tp s_{z-\tp}}{k_2}\right) 
\right. \nonumber\\
&&\times \left.
\left(\sum_{n=1,2,\cdots}\frac{a_{2n}k_2}{E(k_2^2-q_{2n}^2)}\right)
\phi_{2-}
-\frac{c_\tp c_{z-\tp}}{k_1}\left(\sum_{n=1,2,\cdots}
\frac{a_{2n}k}{E(k^2-q_{2n}^2)}\right)\phi_{-}\right\}\;.
\ena
Now, we find
\bea
&&2{\rm Re}[(S_{00})_{e \mu }(S_{01,1s})_{e \mu }^\ast
-(e \iff \mu)]
\nonumber\\
&&\hskip 5mm =-\frac{\Delta m^2_{21}s_{2x}s_{2y}s_{2\tp}s_\delta}{E}
\left(\frac{c_{z-\tp}c_{\tp}}{k_1}
+\frac{s_{z-\tp}s_{\tp}}{k_2}\right)
\nonumber\\
&&\hskip 10mm \times 
\sum_{n=1,2,\cdots}\frac{a_{2n}}{E}\left(
\frac{c_\tp^2 k_1}{k_1^2-q_{2n}^2}
 -\frac{s_\tp^2 k_2}{k_2^2-q_{2n}^2}
\right) \sin \frac{k_1L}2 
\sin \frac{k_2L}2 \sin \frac{kL}2 \;.
\ena
Thus, we obtain the $\sin \delta$ part of T-violation   
as
\bea
\left(\Delta P^{T}_{\nu_e \nu_\mu}\right)_{s_\delta}
&=&-\frac{\Delta m^2_{21}}{E}s_{2x}s_{2y}s_{2\tp}s_\delta 
\left[\frac{c_{\tp}c_{z-\tp}}{k_1}+\frac{s_{\tp}s_{z-\tp}}{k_2}
\right]\sin \left(\frac{k_1L}2\right) 
\sin \left(\frac{k_2L}2\right) \sin \left(\frac{kL}2\right)
\nonumber\\
&&\times \left\{ 1+\sum_{n=1,2,\cdots}\frac{a_{2n}}{E}
\left(\frac{c_\tp^2 k_1}{k_1^2-q_{2n}^2}
 -\frac{s_\tp^2 k_2}{k_2^2-q_{2n}^2}
 +\frac{c_{2\tp}k}{k^2-q_{2n}^2}
\right)\right\}
\;.
\ena
Similarly, we confirmed that the relation 
\bea
\left(\Delta P^{T}_{\nu_\mu  \nu_\tau}\right)_{s_\delta}
=\left( \Delta P^{T}_{\nu_\tau\nu_e}\right)_{s_\delta}
=\left(\Delta P^{T}_{\nu_e \nu_\mu}\right)_{s_\delta}\;,
\ena
holds in the order of 
$(\Delta m_{21}^2L/2E)(\delta a(x)_s L/2E)$.  

\vskip 2mm
\noindent
(b) The contribution from the asymmetric matter fluctuation

The interference term between $\Delta m_{21}^2 L/2E$  and 
the asymmetric matter fluctuation, $\delta a(x)_a$ gives 
the $\cos \delta$ part of  
T-violation. In order to estimate this contribution, 
we consider an extreme case, the linear fluctuation, as a typical 
asymmetric matter profile, 
\bea
\delta a(x)_a=\alpha \bar a\left(\frac{ x-\frac{L}2}{L}\right)\;,
\ena
where $\alpha$ represents the fraction of the asymmetric matter
profile. 

Firstly, we compute $S_{01}S_{1a}^{(1)*}$ term, where 
$S_{1a}^{(1)*}$ is the 1st order term from the asymmetric matter 
profile. For this, we estimate 
${\cal C}_{1}^{(1)}-{\cal D}_{1}^{(1)}$ defined in Eq.(30) 
from $\delta a(x)_a$. We find 
\bea
{\cal C}_1^{(1)}-{\cal D}_1^{(1)}=\frac{\alpha \bar a s_{2\tp}}{2EL}
\left(\frac{L}{2k}\phi_+-i\frac1{k^2}\phi_- \right)\;.
\ena
Now the contribution from the 
interference term between $S_{01}$ and $S_{1a}^{(1)}$ is obtained 
by using Eqs.(21) and (28) as 
\bea
&&2{\rm Re}[(S_{01})_{e \mu}(S_1)_{e \mu}^\ast
-(e \iff \mu)]
\nonumber\\
&&\hskip 5mm =-\frac{\alpha \bar a}{2EL} 
\frac{\Delta m^2_{21}s_{2x}s_{2y}s_{2\tp}c_\delta}{E}
\left(\frac{c_{z-\tp}c_{\tp}}{k_1}
+\frac{s_{z-\tp}s_{\tp}}{k_2}\right)
\nonumber\\
&&\hskip 10mm \times 
\left( \frac{1}{k^2}
-\frac{L}{2k}\cot \frac {kL}2 \right)
\sin \frac{k_1L}2 
\sin \frac{k_2L}2 \sin \frac{kL}2 
 \;.
\ena

Next we consider the $S_{00}S_{01,1a}$ contribution. 
By performing the integrations, we find
\bea
{\cal A}_{01,1a}&=&-{\cal A'}_{01,1a}
\nonumber\\
&=&\frac{\alpha \bar a}{4EL}\frac{\Delta m_{21}^2s_{2x}}{4E}
c_\tp\left( \left(\frac{c_\tp c_{z-\tp}}{k_1}+\frac{s_\tp
s_{z-\tp}}{k_2}
\right)\left( \frac{L}{k_1}\phi_{1+}-i\frac{2}{k_1^2}\phi_{1-}\right)
\right.\nonumber\\
&&\left.-\frac{s_\tp s_{z-\tp}}{k_2}
\left( \frac{L}{k}\phi_{+} -i\frac{2}{k^2}\phi_{-}\right)
\right)
\;, \nonumber\\
{\cal B}_{01,1a}&=&-{\cal B'}_{01,1a}\nonumber\\
&=&\frac{\alpha \bar a}{4EL}\frac{\Delta m_{21}^2s_{2x}}{4E} 
s_\tp\left( -\left(\frac{c_\tp c_{z-\tp}}{k_1}+\frac{s_\tp
s_{z-\tp}}{k_2}
\right)\left( \frac{L}{k_2}\phi_{2+}-i\frac{2}{k_2^2}\phi_{2-}\right)
\right.\nonumber\\
&&\left.+\frac{c_\tp c_{z-\tp}}{k_1}
\left( \frac{L}{k}\phi_{+}-i\frac{2}{k^2}\phi_{-}\right)
\right)
\;.
\ena
Now T-violation   for the 
$\nu_e$ and $\nu_\mu$ oscillation channels is given by
\bea
&&2{\rm Re}[(S_{00})_{e \mu}(S_{01,1a})_{e \mu}^\ast -
(e \iff \mu)]\nonumber\\
&& \hskip 2mm =-\frac{\alpha \bar a}{2EL}
\frac{\Delta m^2_{21}  s_{2x}s_{2y}s_{2\tp}c_\delta}{E}
\left(\frac{c_{z-\tp}c_{\tp}}{k_1}
+\frac{s_{z-\tp}s_{\tp}}{k_2}\right)
\sin \frac{k_1L}2 \sin \frac{k_2L}2 \sin \frac{kL}2 \nonumber\\
&& \hskip 10mm\times \left\{
        -\left(\frac{c_{\tp}^2}{k_1^2}+\frac{s_{\tp}^2}{k_2^2}\right)
    +\frac{L}2 \left(\frac{c_{\tp}^2}{k_1}\cot \frac{k_1L}2
    +\frac{s_{\tp}^2}{k_2} \cot \frac{k_2L}2\right)
   \right\}
\;.
\ena
By adding these contributions, we obtain the asymmetric matter effect 
to T-violation,  
\bea
\left(\Delta P^T_{\nu_e\nu_\mu}\right)_{c_\delta}
&=&-\frac{\alpha \bar a}{2EL} 
\frac{\Delta m^2_{21} s_{2x}s_{2y}s_{2\tp}c_\delta}{E}
\left(\frac{c_{\tp}c_{z-\tp}}{k_1}+\frac{s_{\tp}s_{z-\tp}}{k_2}\right)
\sin \frac{k_1L}2 
\sin \frac{k_2L}2 \sin \frac{kL}2 \nonumber\\
&\times&  \left\{
        \left(\frac{1}{k^2}-\frac{c_{\tp}^2}{k_1^2}
        -\frac{s_{\tp}^2}{k_2^2}\right)
-\frac{L}2 \left(\frac{1}{k} \cot\frac{kL}{2}
-\frac{c_{\tp}^2}{k_1}\cot \frac{k_1L}{2}
-\frac{s_{\tp}^2}{k_2}\cot \frac{k_2 L}{2}\right)
\right\}\;.
\nonumber\\
\ena
Similarly, we confirmed
\bea
\left(\Delta P^T_{\nu_e\nu_\mu}\right )_{c_\delta}
=\left(\Delta P^T_{\nu_\mu \nu_\tau}\right )_{c_\delta}=
\left(\Delta P^T_{\nu_\tau \nu_e}\right )_{c_\delta}\;.
\ena

From Eqs.(58) and (65), we find that T-violation   is 
independent of flavor in the 2nd order perturbation.

\section{Numerical analysis}

In order to analyze the matter effect, we use the PREM 
profile for the symmetric matter density distribution.

\vskip 2mm
\noindent
(a) $L=3000$km 

\vskip 2mm
\noindent
(a-1) The T-violation (intrinsic) asymmetry 

Firstly, we estimate how large the T-violation asymmetry is. 
In Fig.1, we plotted 
$(\Delta P^T_{\nu_e\nu_\mu})_{s_\delta}
/2P(\nu_e \to \nu_\mu)$ as a function of the neutrino 
energy $E$, with $\delta=\pi/4$, $s_{2x}=s_{2y}=1$, 
$s_z=0.1$. Here, we used our formula in Eq.(57) for 
$(\Delta P^T_{\nu_e\nu_\mu})_{s_\delta}$ and the 0th order term 
for $P(\nu_e \to \nu_\mu)$. For the symmetric matter profile, 
we use $\bar \rho=3.31602{\rm g}/{\rm cm}^3$ 
and the Fourier coefficients, 
$a_2$,  $a_4$,  $a_6$ and $a_8$ are $-$0.045, $-$0.048, $-$0.047 and 
$-$0.044${\rm g}/{\rm cm}^3$ respectively, 
which are derived by Ota and Sato[9] 
from PREM. We neglect the $n\ge 10$ terms. As we can see 
from this figure, we can expect about $4 \sim 10\%$ effect for 
$E>5$GeV.  

\vskip 2mm
\noindent
(a-2) The matter-modified T-violation  
in the symmetric matter profile

In the 2nd order perturbation, the symmetric matter gives the 
contribution to the $\sin \delta$ of T-violation  and 
the combined formula is given in Eq.(57). 

In Fig.2, the comparison between our result and the vacuum case 
of T-violation,
\bea
\left(\Delta P^T_{\nu_e\nu_\mu}\right)_{\rm vacuum}
=-s_{2x}s_{2y}s_{2z}c_z s_\delta
\sin \frac{\Delta m_{21}^2 L}{4E}
\sin \frac{\Delta m_{31}^2 L}{4E} \sin \frac{\Delta m_{32}^2 L}{4E}
\;.
\ena
is made, for (a) 3GeV$<E<$30GeV and 
(b) 1GeV$<E<$3GeV. The values of parameters are the same as those 
in Fig.1. The solid line shows the vacuum 
case and the dash-dotted line shows the matter-modified 
T-violation with the addition of the 1st and the 2nd order terms. The 
2nd order contribution is shown by the dotted line, but it is hard 
to see because it is almost zero in this scale. That is, the 
2nd order term is negligibly small and we can safely use the 
formula given by taking the average density. There is the matter 
enhancement around $E=6$GeV which is consistent with the discussion 
by Parke and Weiler[7]. 

\vskip 2mm
\noindent
(a-3) The matter-modified T-violation  
in the asymmetric matter profile

In Fig.3, we plotted $(\Delta P)_{c_\delta}$, the fake contribution to 
T-violation from the asymmetric matter fluctuation, 
for (a) 3GeV$<E<$30GeV and (b) 1GeV$<E<$3GeV. The values of parameters 
are the same as those in Fig.1. In addition, we considered 
the $10\%$ asymmetric matter fluctuation, $\alpha=0.1$. 
We observe that 
for $E>$5GeV, the contribution is much less than 1$\%$ of the 
intrinsic T-violation and for 
1GeV$<E<$5GeV, the contribution is at most $3\%$. Thus we conclude 
that the asymmetric matter contribution is negligibly small 
for most energies. The small contribution for T-violation  
asymmetry from the asymmetric matter may be understood by 
observing that the content of the curly parenthesis in Eq.(64)
vanishes 
when $|k_1L|<<1$, $|k_2L|<<1$ and $|kL|<<1$.

We expect that the linear shape for the asymmetric matter is 
the biggest deviation from the symmetric matter. In order to 
see the shape dependence of the contribution, we consider 
the cosine shapes and discuss which cosine shape in the 
Fourier series will give 
the largest contribution to T-violation. 
That is, we consider the asymmetric matter fluctuation by 
\bea
\delta a(x)= -\frac{4}{\pi^2}\alpha_n\bar a \cos(q_{2n+1}x)\;,
\ena  
where $q_{2n+1}=(2n+1)\pi/L$. The coefficient, $4/\pi^2$ is 
attached by normalizing to the linear shape, 
$(x-L/2)/L=-4/\pi^2\sum_{n=0,1,\cdots}
(2n+1)^{-2}\cos (2n+1)\pi x/L$. 
Now we compare the contribution for various $n$ by taking 
$\alpha_n=0.1$. 
We find 
\bea
\left(\Delta P^T_{\nu_e\nu_\mu}\right)_{n}
&=&\frac{4\alpha_n\bar a \Delta m^2_{21} s_{2x}s_{2y}s_{2\tp}c_\delta}
{2\pi^2E^2}
\left(\frac{c_{\tp}c_{z-\tp}}{k_1}+\frac{s_{\tp}s_{z-\tp}}{k_2}\right)
\sin \frac{k_1L}2 
\sin \frac{k_2L}2 \sin \frac{kL}2
\nonumber\\
&\times& 
\left\{ \frac{k}{k^2-q_{2n+1}^2}\cot \frac{kL}2
-\frac{c_{\tp}^2k_1}{k_1^2-q_{2n+1}^2}\cot \frac{k_1 L}2
-\frac{s_{\tp}^2k_2}{k_2^2-q_{2n+1}^2}\cot \frac{k_2 L}2\right\}
\;.\nonumber\\
\ena

Firstly, we checked that the $n=0$ case with $\alpha_0=0.1$ 
agrees with the linear shape case with $\alpha=0.1$. We  
confirmed that the difference around edges does not make any 
difference and the agreement is quite good. Here, we compare 
$n=1,2,3$ cases with $n=0$ case in Fig.4, with the same values 
of parameters as in Fig.1.  
That is, we plotted $(\Delta P)_{n}/(\Delta P)_{n=0}$ for 
$n=1,2,3$ with $\alpha_n=0.1$  for $E>5$GeV. We found that 
as $n$ becomes larger, the contribution becomes smaller.  
The $n=1,2$ and 3 cases give about 1/2, 1/5 and 1/10 times smaller 
than the $n=0$ case, respectively. Thus, we conclude that the 
linear shape case gives the largest contribution to T-violation. 

\vskip 2mm
\noindent
(b) $L=7332$km case

We use $\bar \rho=4.21498{\rm g}/{\rm cm}^3$ 
and the Fourier coefficients, $a_2$,  $a_4$,  $a_6$ and $a_8$ 
are $-$0.31, $-$0.13, $-$0.035 and 0.01${\rm g}/{\rm cm}^3$,
respectively[9]. 
In Fig.5, our formula for the $\sin \delta$ part in Eq.(57) is plotted 
in comparison with the vacuum contribution. The 
solid line shows the vacuum case, while the 1st and 
the 2nd order terms are shown by the dotted and the dashed
lines for 5GeV$<E<$20GeV, respectively. We observe that the 2nd order 
term is comparable to the 1st order term, and moreover 
they cancel each other.  The net contribution shown by the 
dash-dotted line is quite small. In Fig.6, the asymmetric matter 
contribution ($\cos \delta$ part) is shown in the solid line, 
in comparison with the $\sin \delta$ part. As we see from the 
figure, the $\cos \delta$ part is as comparable as the 
$\sin \delta$ due to the severe cancellation between the 
1st and the 2nd order terms, though it is 
much smaller than the vacuum case.

From the above analysis, we observe the followings for $L=7332$km:  

(1) We may need to calculate the 3rd order contribution to obtain 
the accurate formula for the $\sin \delta$ part in order to check that 
T-violation   is really as small as we obtained.  

(2) The matter fluctuation is no more neglected, because  
$\delta a(x) L/2E\sim 1/3$ and the convergence of the perturbation is 
not fast. This is because the PREM distribution 
has the $\sin \pi x/L$ like structure for $L=7332$km and thus the 
symmetric matter profile can not be approximated by the 
constant (average) density distribution. 
Therefore, if the PREM distribution is correct, the fluctuation 
from the average medium really 
gives the important contribution for a small quantity such as 
T-violation. 

\section{Summary}

In this paper,  we gave the analytical expressions of 
the contributions from the symmetric and asymmetric matter 
density fluctuations to 
T-violation   in the 2nd order perturbation with 
respect to $\delta m_{21}^2L/2E$ and $\delta a(x)L/2E$. 
We found that the contribution to T-violation arises 
from the interference between $\delta m_{21}^2L/2E$ 
and $\delta a(x)L/2E$. 
The matter fluctuation only does not give any contribution 
to T-violation, which we confirmed by calculating 
up to the 3rd order terms of $\delta a(x)L/2E$. 
The symmetric and the asymmetric 
matter fluctuations give the effect to the $\sin \delta$ 
and the $\cos \delta$ terms, respectively.  
These analytic  formula are quite accurate for the distance 
less than $L=3000$km and can be used to discuss 
T-violation   analytically. 

By analyzing these formula numerically, we found the following results: 
Both the contributions from the symmetric and asymmetric 
matter density fluctuations of the order of less than 10$\%$ to
T-violation are small for $L= 3000$km or a shorter length. 
The use of the average matter density is sufficient 
for the practical use. 
Therefore, we conclude that the observation of
T-violation  with $L=3000$km or a shorter length will give 
a quite clear method to determine the CP violation angle, $\delta$. 

For the $L=7332$km, the situation changes drastically. We found 
that the 2nd order term from the symmetric matter fluctuation 
of the order of less than 10$\%$
is comparable to the 1st order term which includes the effect 
from the constant (average) matter density, 
and moreover they cancel each other. The 
net contribution to the $\sin \delta$ term becomes as small as 
the contribution from the asymmetric matter fluctuation. 
This shows that the symmetric matter fluctuation becomes 
important for $L=7332$km. This is because the PREM distribution 
has the $\sin \pi x/L$ like structure for $L=7332$km and it is 
not approximated by the constant (average) distribution. 
Therefore, if the PREM distribution is correct, 
the constant (average) density approximation does not work for 
$L=7332$km and a longer distance. For $L=7332$km, we 
may need to evaluate the 3rd order term to obtain the accurate 
formula, which is now under the study.

\vskip 5mm
{\Huge Acknowledgment}
This work is supported in part by
the Japanese Grant-in-Aid for Scientific Research of
Ministry of Education, Science, Sports and Culture,
No.12047218.

\begin{figure}[pht]
\epsfxsize=16cm
\centerline{\epsfbox{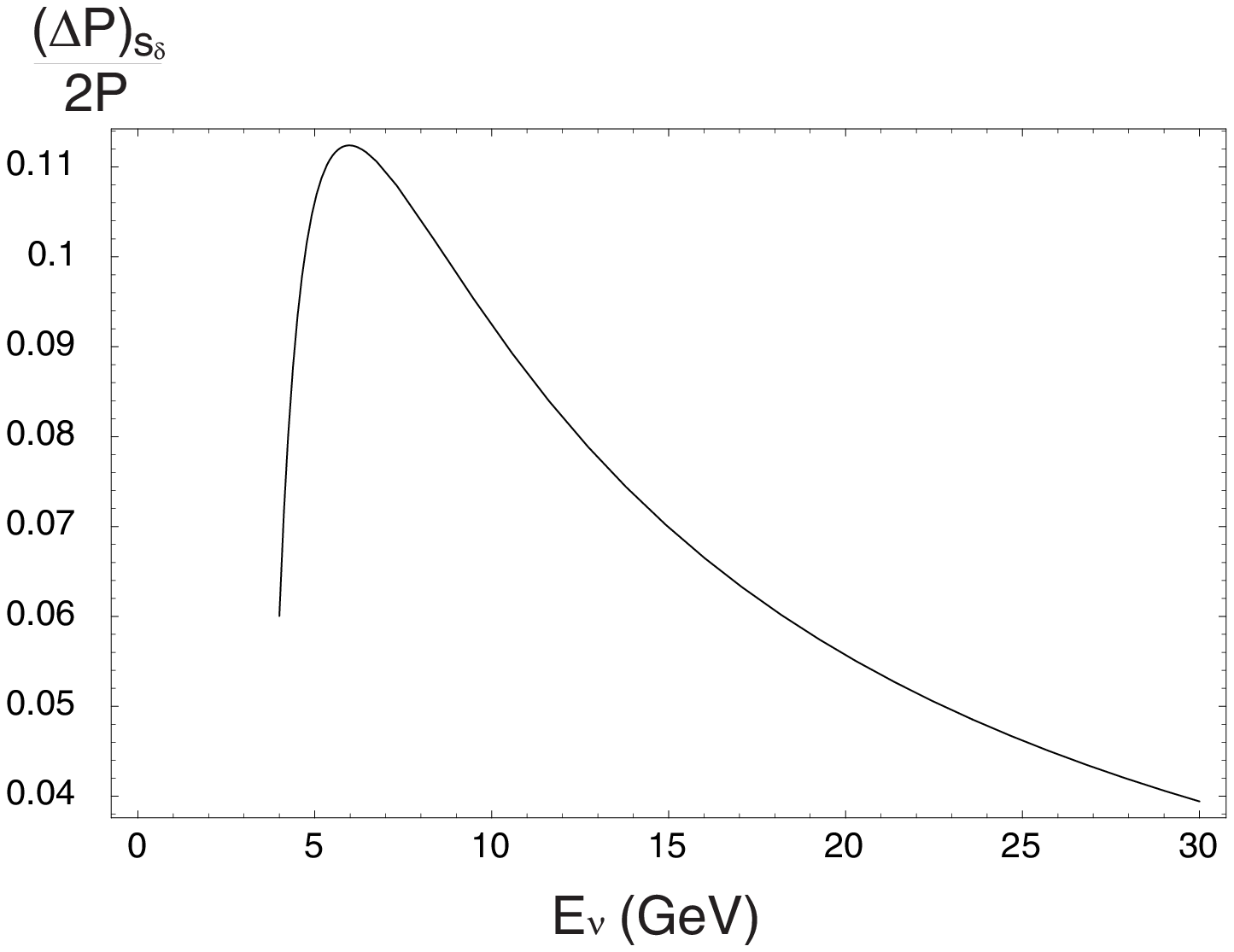}}
\caption{The energy dependence of the T-violation asymmetry 
for the $\nu_e$ to $\nu_\mu$ channel for $L=3000$km. 
$(\Delta P)_{s_\delta}$ represents T-violation 
with the symmetric matter profile which is proportional 
to $\sin \delta$ as given in Eq.(57), and  
$P=P(\nu_e \to \nu_\mu)$ in oscillation probability in the 0-th order. 
For the symmetric matter profile, the PREM distribution is used. 
In this plot, we use $\sin 2\theta_x=\sin 2\theta_y=1$, 
$\sin \theta_z=0.1$,  $\Delta m_{21}^2=5\cdot 10^{-5}{\rm eV}^2$,  
$\Delta m_{31}^2=3\cdot 10^{-3}{\rm eV}^2$ and $\delta=\pi/4$.}
\end{figure}

\begin{figure}[pht]
\epsfxsize=12.4cm
\centerline{\epsfbox{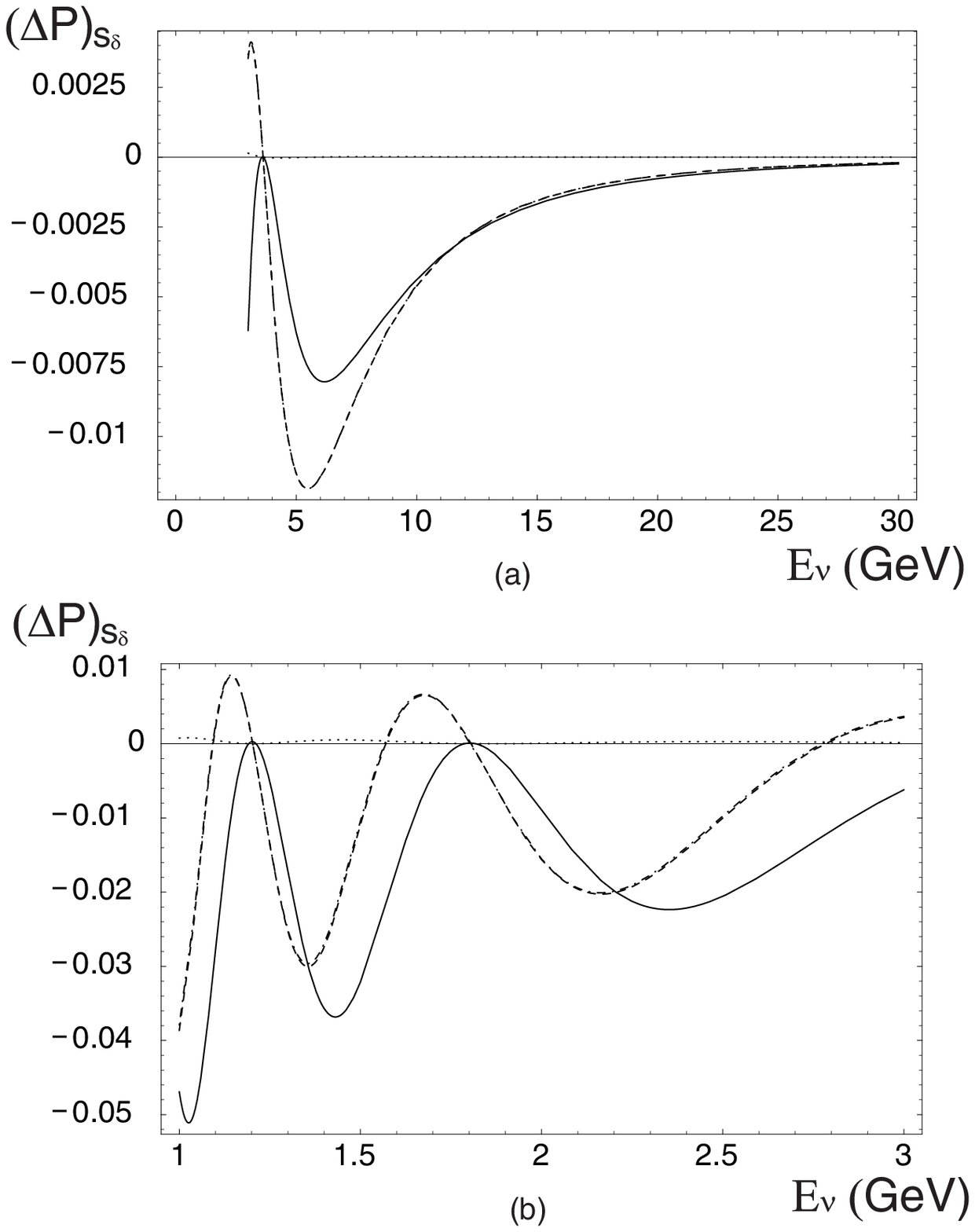}}
\caption{The energy dependence of T-violation   
$(\Delta P)_{s_\delta}$ (the dash-dotted line) 
in comparison with the vacuum case (solid line) with 
$L=3000$km for (a) 3GeV$<E<$30GeV and (b) 1GeV$<E<$3GeV. 
We used the PREM distribution 
for the symmetric matter profile. The dashed line shows 
the 1st order term, which include the constant (average) matter 
contribution, while the contribution from the 2nd order symmetric 
matter is indicated by the dotted line.
The 2nd order term gives negligible contribution and it is hard 
to see in this scale. The difference between $(\Delta P)_{s_\delta}$ 
and the vacuum case around $E=$6GeV is the matter effect 
due to the constant (average) density.}
\end{figure}

\begin{figure}[pht]
\epsfxsize=13cm
\centerline{\epsfbox{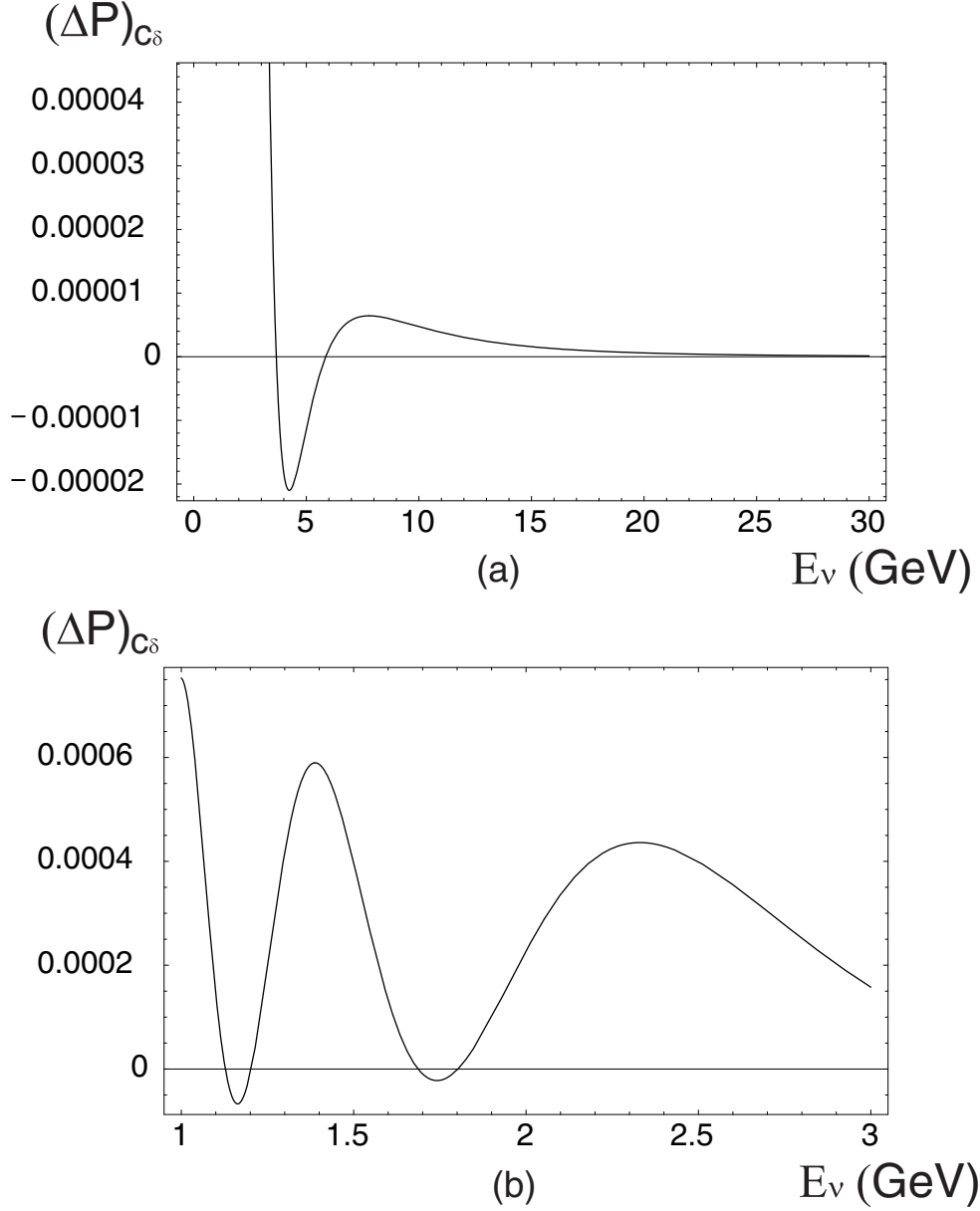}}
\caption{The energy dependence of the asymmetric matter 
contribution (a linear shape) to T-violation, which is 
proportional to $\cos \delta$, $(\Delta P)_{c_\delta}$ 
with $L=3000$km for (a) 3GeV$<E<$30GeV and (b) 1GeV$<E<$3GeV. 
We assumed the $10\%$ asymmetry of the average 
density. The oscillation parameters are the same as those in Fig.1. 
By comparing this with Fig.2, we see the asymmetric matter 
contribution is negligibly small for $E>5$GeV and 
about $3\%$ for $1{\rm GeV}<E<5$GeV.}
\end{figure}

\begin{figure}[pht]
\epsfxsize=16cm
\centerline{\epsfbox{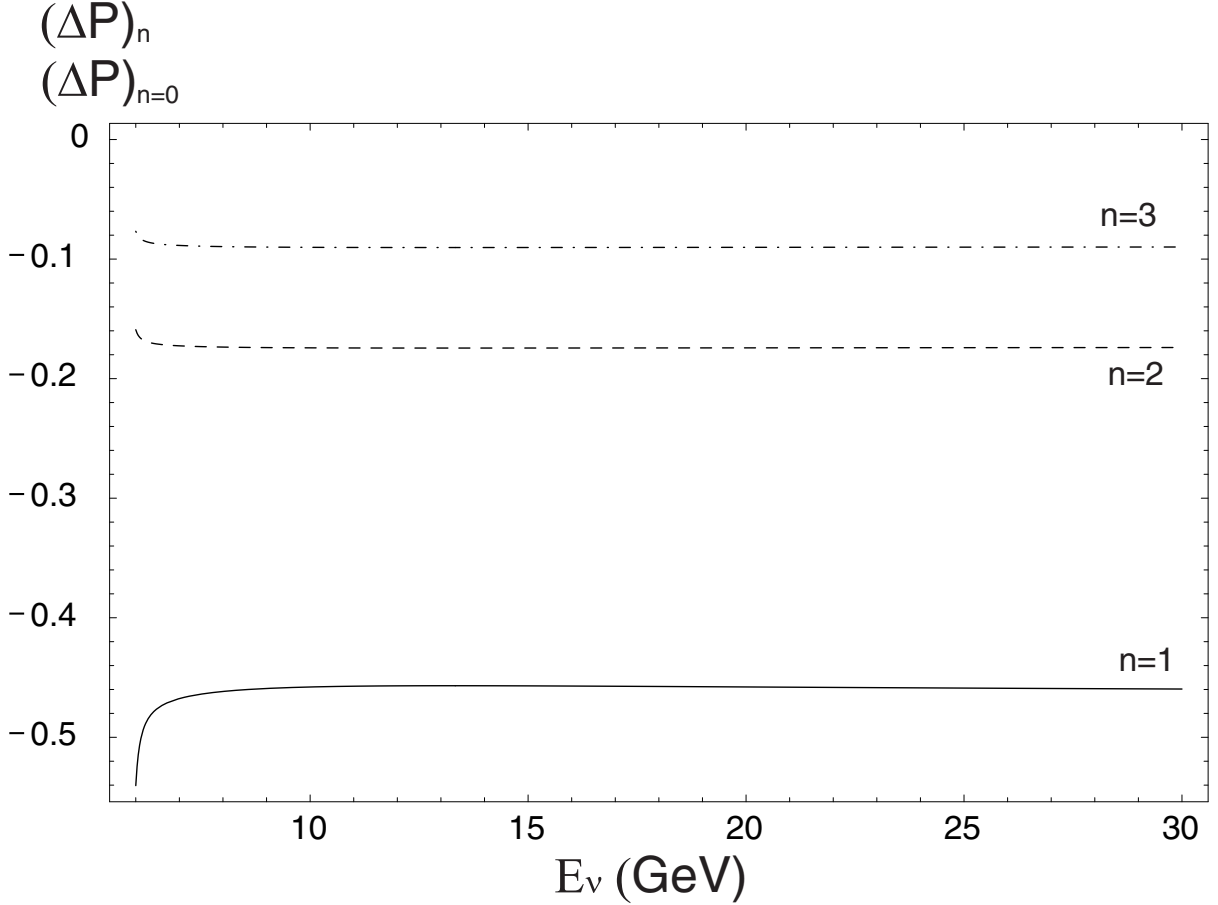}}
\caption{The shape dependence of T-violation   
for the asymmetric matter. Shapes are expressed by 
$\cos (2n+1)\pi x/L$. We confirmed that T-violation  
from the $n=0$ case agrees with that from the 
linear shape case if we choose an appropriate normalization. 
We plotted the ratios of the contributions from $n=1,2,3$ and 
$n=0$. We observe that 
the contribution from the higher $n$ is suppressed by 
about 1/2, 1/5 and 1/10 in comparison with the $n=0$ case. 
This shows that the linear shape or $n=0$ case will give 
the biggest contribution to T-violation.
}
\end{figure}

\begin{figure}[pht]
\epsfxsize=16cm
\centerline{\epsfbox{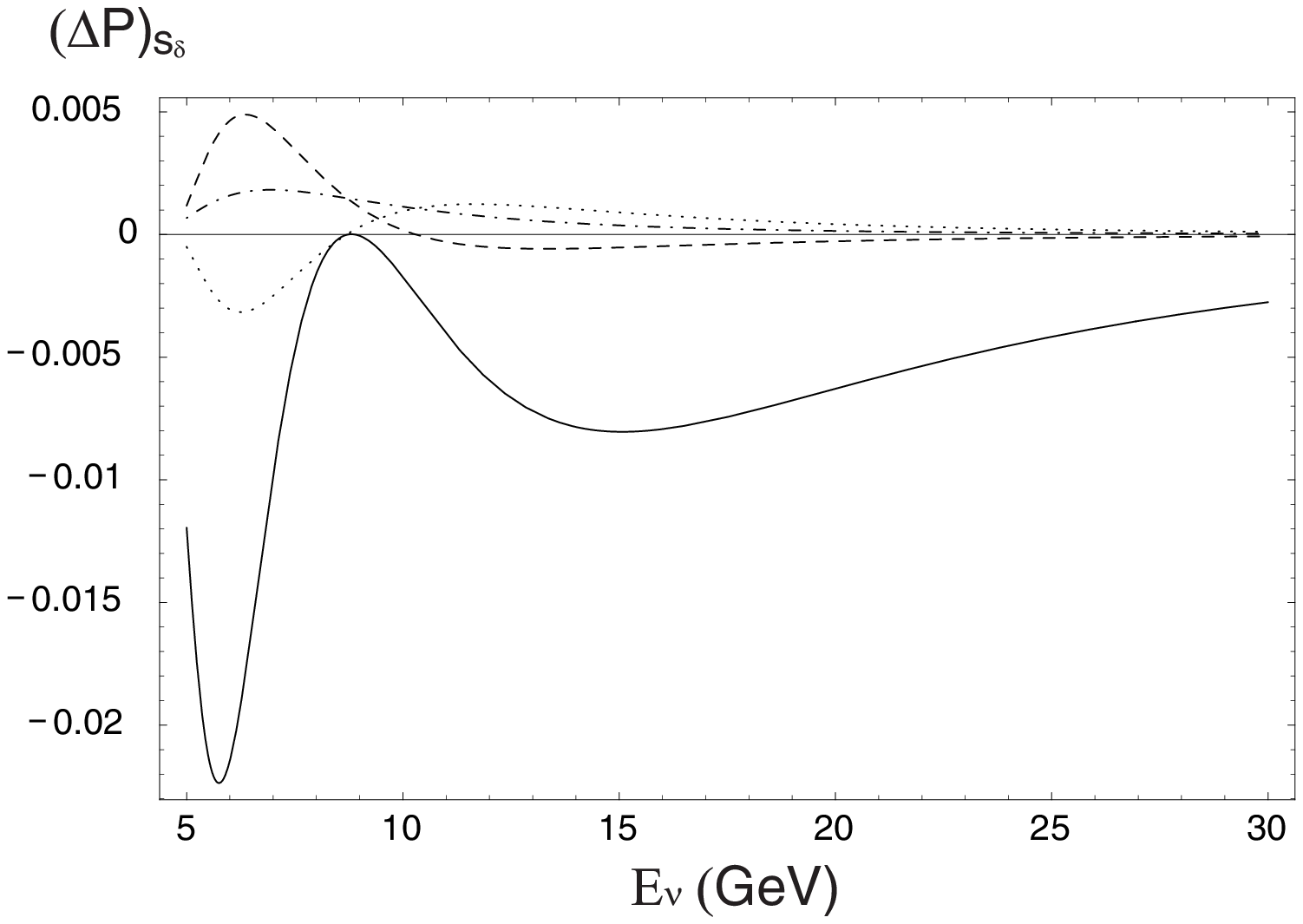}}
\caption{The energy dependence of T-violation   
from the symmetric matter, $(\Delta P)_{s_\delta}$ (the dash-dotted
line) 
in comparison with the vacuum case (solid line) with 
$L=7332$km for 5GeV$<E<$30GeV.  We used the PREM distribution 
for the symmetric matter. The dashed line shows 
the 1st order term, which include the constant (average) matter 
contribution, while the contribution from the 2nd order symmetric 
matter is indicated by the dotted line. The oscillation parameters 
are the same as those in Fig.1.  There is severe 
cancellation between the 1st and the 2nd order terms, and 
the net contribution becomes much smaller. This shows that the 
matter fluctuation gives a sizable effect to the 
$\sin \delta$ term and the constant (average) 
approximation for the symmetric matter does not give a 
good approximation.}
\end{figure}

\begin{figure}[pht]
\epsfxsize=16cm
\centerline{\epsfbox{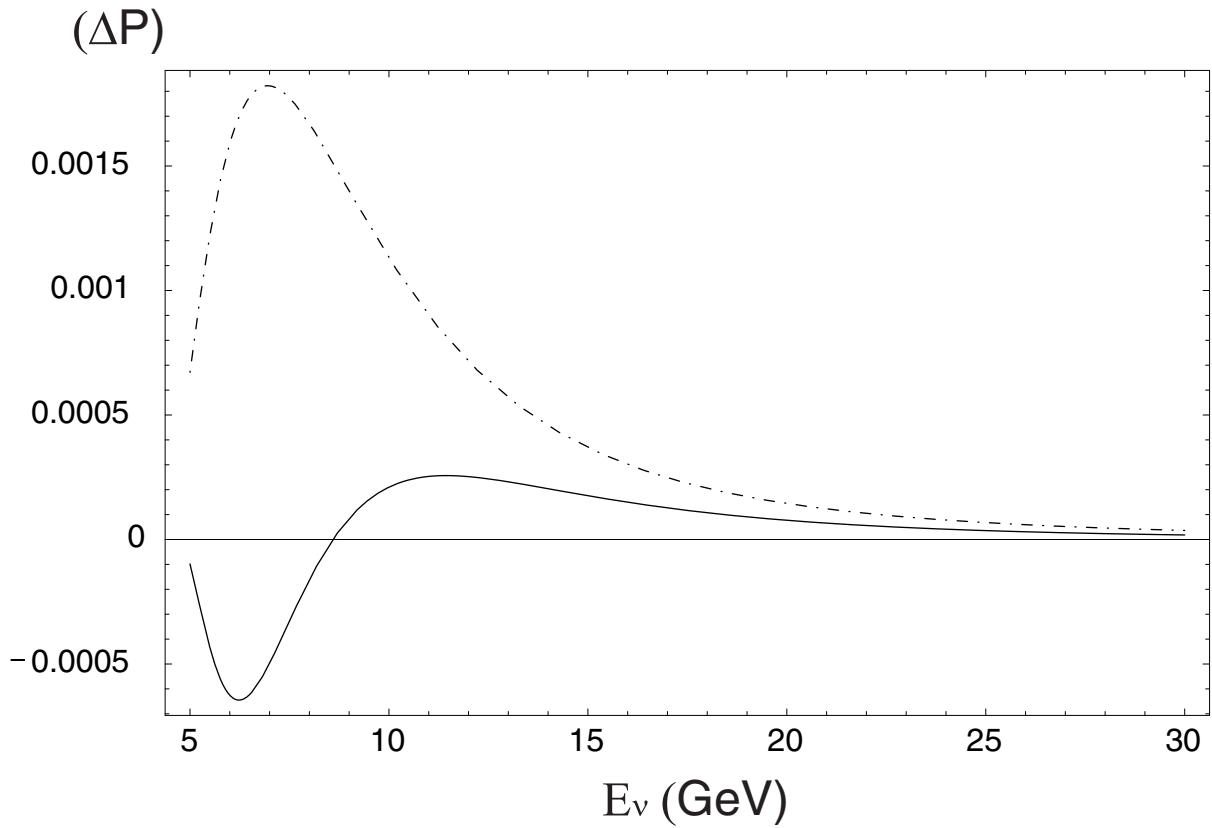}}
\caption{The comparison between T-violation   
from the symmetric matter ($(\Delta P)_{s_\delta}$, 
the dash-dotted line) and from the asymmetric matter 
($(\Delta P)_{c_\delta}$, the solid line) for $L=7332$km. 
We used the PREM for the symmetric matter and assumed 
the asymmetry of $10\%$ in comparison with the average density. 
The oscillation parameters are the same as those in Fig.1.  
The $\sin \delta$ part is not the dominant term in contrast to 
$L=3000$km case. This is due to the severe cancellation 
between the 1st term (including the effect from the 
constant (average) 
mater) and the 2nd order term from the symmetric matter 
fluctuation. }
\end{figure}

\end{document}